\documentclass
[12pt,tightenlines,eqsecnum,floats,aps,amsmath,amssymb,tightenlines,nofootinbib,prd,shownopacs,notitlepage]{revtex4}

\UseRawInputEncoding

\usepackage{graphicx, wrapfig}
\usepackage{amssymb}
\usepackage{color}
\usepackage{mathrsfs}
\usepackage{amsmath}
\usepackage{subfigure}
\usepackage{afterpage}

\usepackage{epstopdf}

\DeclareGraphicsExtensions{.pdf,.png,.jpg,.jpeg,.eps}

\def\f{\frac}

\def\ig{\includegraphics}

\def\lp{\ell_{\rm Pl}}
\def\t{\tilde}
\def\h{\hat}

\def\EE {\rm EEs\,}

\def\R{\mathcal R }

\def\rmd{{\rm d}}

\def\L{{\rm L}}

\usepackage{enumerate}

\def\dd{{\rm d}}

\def\dd{\rm d}

\def\kz{k_{\circ}}

\def\Az{\mathring{A}}
\def\be{\nopagebreak[3]\begin{equation}}
\def\ee{\end{equation}}
\def\ba{\nopagebreak[3]\begin{eqnarray}}
\def\ea{\end{eqnarray}}

\newcommand{\bfig}{\begin{figure}[h]}
\newcommand{\efig}{\end{figure}}

\newcommand{\bmult}{\nopagebreak[3]\begin{multline}}
\newcommand{\emult}{\end{multline}}

\newcommand{\fref}[1]{Fig.\,\ref{#1}}

\def\rcurv{r_{\rm curv}}

\def\lcdm{$\Lambda{\rm CDM}~$}
\def\mpc{{\rm Mpc^{-1}}}

\def\planck{{PLANCK} }
\def\klqc{k_{\rm LQC}}

\def\RB{{\mathfrak R}_{\rm B}}

\def\TT{\rm TT}
\def\TE{\rm TE}
\def\EE{\rm EE}
\def\BB{\rm BB}
\def\SA{the {\it SA\,\,}}

\def\H{\mathcal{H}}
\def\Hlqc{\mathcal{H}_{\rm LQC}}
\def\Q{\mathcal{Q}}
\def\U{{\mathcal{U}}}

\begin{document}

\title{
Cosmic tango between the very small and the very large:
Addressing CMB anomalies through Loop Quantum Cosmology}

\author{Abhay Ashtekar${}^1$\,}
\email{ashtekar.gravity@gmail.com}
\author{Brajesh Gupt${}^{2,1}$\, }
\email{brajeshgupt@gmail.com}
\author{V. Sreenath${}^3$\,}
\email{sreenath@nitk.edu.in}
\affiliation {
${}^1$ Institute for Gravitation and the Cosmos \& Physics Department,\\ 
The Pennsylvania State University, University Park, PA 16801\\
${}^2$ Texas Advanced Computing Center,
The University of Texas at Austin, Austin, TX 78758, USA\\
${}^3$ Department of Physics, National Institute of Technology Karnataka, Surathkal, India 575025}

\begin{abstract}
While the standard, six-parameter, spatially flat \lcdm model has been highly successful, certain anomalies in the cosmic microwave background bring out a tension between this model and observations. The statistical significance of any one anomaly is small. However, taken together, the presence of two or more of them imply that according to standard inflationary theories we live in quite an exceptional universe.  We revisit the analysis of the PLANCK collaboration using loop quantum cosmology, where an unforeseen interplay between the ultraviolet and the infrared makes the \emph{primordial} power spectrum scale dependent at very small $k$. Consequently, we are led to a somewhat different \lcdm universe in which anomalies associated with large scale power suppression and the lensing amplitude are both alleviated. The analysis also leads to new predictions for future observations. This article is addressed both to cosmology and LQG communities, and we have attempted to make it self-contained.

\end{abstract}

\maketitle

\section{Introduction}
\label{s1}

The quantum geometry effects underlying loop quantum gravity (LQG) lead to a natural resolution of the big bang singularity (see, e.g., \cite{aapsrev,30years:IAPS} for reviews). Therefore, one can hope to meaningfully extend the standard inflationary paradigm to the Planck regime. Over the past decade, several closely related approaches have been used to carry out this task, leading to a striking interplay between theory and observations (see, in particular, \cite{aan1,aan2,aan3,madrid,bcgmrev,lcbg,aaab,agullomorris,agulloassym,ag2,ag3,menaetal,abs,sab,bjmm,aos1,aos2,aks1,aks}). In this article we will focus on the recent results that shed new light on the anomalous features seen in the cosmic microwave background (CMB). Specifically, we will show that in our approach two of the anomalies seen in the CMB can be accounted for using the pre-inflationary dynamics of loop quantum cosmology (LQC). This phase of dynamics alters the quantum state of cosmological perturbations at the onset of the (relevant part of the) slow roll, leading to revised values of the six parameters that characterize the \lcdm universe. The revision alleviates the tension due to two anomalies that have received considerable attention, while leaving the successes of standard inflation intact. Main results were reported in \cite{agjs}. The purpose of this paper is to provide details and also present some supplementary material to put the results in a broader context. These results illustrate that LQC has matured sufficiently to lead to testable predictions.

The paper is addressed both to the LQG community and cosmologists. For the benefit of the LQG community, that primarily focuses on mathematical physics, we have included a discussion of the interplay between theory and observations that leads to the six parameter \lcdm cosmological model. We will summarize the underlying procedure and point out certain subtleties in data analysis. 
For cosmologists, we will summarize the key features of LQC that lead to new observable predictions. Specifically we will explain how the quantum geometry effects \emph{in the ultraviolet}, that lead to the singularity resolution, have unforeseen and interesting consequences on the dynamics of cosmological perturbations \emph{in the infrared}. It is this `cosmic tango' between the very small and the very large that alleviates anomalies. Overall, in terms of conceptual flow, we have attempted to make this paper self-contained. In particular, within the page limits of this special issue, we clarify apparently conflicting statements in the LQC literature. In order to make the material accessible to both communities, we will have to briefly review ideas and results that are likely to be well-known in one community but not the other. 

The two anomalies we focus on arise as follows. Motivated in large part by inflationary scenarios, the CMB analysis generally begins by assuming that the \emph{primordial} scalar power spectrum has a nearly scale invariant form, characterized by just two parameters, the scalar perturbation amplitude $A_s$ and the scalar spectral index $n_s$. We will refer to this form as the \emph{standard ansatz (SA)}. $A_s,\, n_s$ and 4 other parameters (discussed in section \ref{s2.1}) characterize a specific \lcdm universe. Given these six parameters one can evolve the primordial perturbations using known astrophysics and predict the observable power spectra. By varying the values of the 6 parameters, and confronting the theoretical prediction with observations, one finds the posterior probability distributions of the six cosmological parameters. By and large the CMB observations can be well explained using the \lcdm universe determined by the marginalized mean values of these parameters. However, one also finds some anomalous features. The first is power suppression at large angular scales: the observed power in the temperature-temperature (TT) spectrum is suppressed for $\ell \lesssim 30$ in the spherical harmonic decomposition, relative to what the theory predicts. The second anomaly we focus on is associated with the so-called lensing amplitude, $A_{L}$, associated with gravitational lensing that the CMB photons experience as they propagate from the surface of last scattering to us. The \lcdm cosmology based on \SA assumes $A_{L} =1$ while, when it is allowed to vary, $A_L$ prefers a value larger than unity.   This tension hints at an internal inconsistency. To alleviate it, one can introduce a positive spatial curvature but then there are inconsistencies with the low $z$ measurements, prompting a recent suggestion \cite{silketal} that this anomaly gives rise  to a ``possible crisis in cosmology.'' 

As we will see, both the anomalies are simultaneously alleviated in our approach. The key new element is the following: Pre-inflationary dynamics of LQC leads to a \emph{primordial} power spectrum that differs from {\it SA}, but only at large angular scales. While it continues to be nearly scale invariant --and essentially indistinguishable from the one given by \SA--  for $\ell \gtrsim 30$, there is a specific power suppression for $\ell \lesssim 30$. As a result, the best-fit values of the six cosmological parameters change. Interestingly, the change in 5 of the 6 parameters is extremely small, $\lesssim 0.4\%$. But the value of the 6th parameter --the optical depth $\tau$-- is increased by $\sim 9.8\% $! We will see that this change then leads to the alleviation of the tension between observations and theoretical predictions based on the {\it SA}. Note that in spite of this significant change in the value of $\tau$, LQC leaves the highly successful predictions of standard inflation at small angular scales unaffected. In particular, all the finer features of various power spectra predicted by standard inflation for $\ell >  30$ --where the observational error bars are small-- are present also in the LQC prediction. Thus the LQC analysis provides an explicit example supporting a conclusion of \cite{inflation-martin} that trans-Planckian effects are not a ``threat to inflation".

The paper is organized as follows. Section \ref{s2} summarizes the procedure used in observational cosmology to arrive at the 6 parameter \lcdm model and explains the two anomalies and their significance in greater detail. Section \ref{s3} summarizes the basic results from LQC that are used in the subsequent analysis. In particular, we explain the origin of the surprising interplay between the ultraviolet and the infrared that is a rather robust feature of the LQC approaches.  The main results are presented in Section \ref{s4}. They include a discussion of: (i) the LQC corrected \emph{primordial} power spectrum for scalar perturbations;\, (ii) the TT, temperature-electric polarization (TE), the electric polarization (EE), and the lensing potential ($\phi\phi$) power spectra we predict, and comparisons with those obtained using \SA as well as with the observed power spectra reported by the PLANCK team in their final analysis \cite{planck5}. As usual these power spectra are presented in terms of the spherical harmonic components $C_\ell$ of the respective correlation functions;\,
(iii) the TT correlation function $C(\theta)$ predicted by LQC and its comparison with the 
prediction of \SA as well as PLANCK observations;\,
(iv) the $A_L$ versus $\tau$  plots that show that the observed values fall in the $1\sigma$ contour in LQC, but not if one uses the {\it SA}; and,\,
(v) the power spectrum for BB polarization predicted by LQC  and its comparison with that predicted by the {\it SA}.\,
The detailed calculations underlying these plots were performed using the Starobinsky and quadratic potentials. The first is preferred phenomenologically while the second has been used often because of its simplicity. Close agreement between the two sets of results is an indication of robustness of the LQC results. In Section \ref{s5} we summarize the main results and put them in a broader context. 

So far the discrepancy  between the results of the SHOES team \cite{SHOES} and CMB measurements \cite{planck6} associated with the value of the Hubble parameter has not been systematically addressed in LQC. This is in large part because it is not yet clear whether there is a definitive tension, or if the discrepancy is primarily due to systematic calibration offsets \cite{ge}. Observations may decide on this issue in the near future.

\section{Theoretical predictions and PLANCK observations}
\label{s2}

This section is addressed primarily to the LQG community. In \ref{s2.1} we summarize the procedure used by the PLANCK team to arrive at the six parameter \lcdm model and in \ref{s2.2} we explain the power suppression and the lensing amplitude anomalies in a bit more detail.

\subsection{The 6 parameter $\Lambda$CDM model}
\label{s2.1}

The six parameters that characterize the \lcdm universe can be neatly divided in three groups. The first two parameters --the amplitude $A_s$ and the spectral index $n_s$ for scalar modes--
feature in \SA for the primordial power spectrum: 
\be \label{SA} 
\mathcal{P}_{\mathcal{R}}(k) = A_{s}\, 
\left(\f{k}{k_{\star}}\right)^{n_{s}-1}\!\!\!. \ee 
Here $k$ is the wave number in the Fourier decomposition and $k_\star$ is a pivot scale (set to $k_\star\, =\, 0.002 {\rm Mpc}^{-1}$ in the WMAP analysis and $k_\star \!=\! 0.05 {\rm Mpc}^{-1}$ in the PLANCK analysis). If we had $n_s\! =\!1$, the primordial spectrum would be scale invariant, i.e., it would be independent of the wave number $k$ of the cosmological perturbation. If $n_s$ is less than $1$ (as observations imply) then there is more power at small $k$, i.e., the power spectrum has a red tilt. (One can also consider the possibility of a running $n_s$, where it has a $k$ dependence but we will not need this generality.) The second set of parameters, the baryonic and cold matter densities $\Omega_b h^2$ and $\Omega_c h^2$, are important for the propagation of cosmological perturbations starting from the end of inflation. The last group of parameters are $100 \theta_{MC}$, that characterizes the angular scale of acoustic oscillations, and the optical depth $\tau$ that characterizes the reionization epoch. These two parameters govern the propagation of perturbations from the last scattering surface to now. Thus, given these six parameters one can use the known astrophysics to propagate the cosmological perturbations starting from the end of inflation, providing us with the theoretical predictions for power spectra we should observe now.

More precisely, for each choice of the six parameters, one can calculate 4 correlation  functions $C^{\TT}_\ell,\, C^{\TE}_\ell,\, C^{\EE}_\ell,\, C^{\phi\phi}_\ell$ that are the spherical harmonic decompositions of the corresponding correlation function $C^{\rm XY}(\theta)$ (where $\theta$ characterizes the angular separation of two points in the sky). These correlation functions can be measured and compared with the theoretical predictions, expressed as functions of the six parameters. The statistical analysis, usually done by employing Markov-Chain Monte Carlo method, then leads to the posterior probability distribution for the six parameters. In particular, the maximum likelihood value of the marginalized probability distributions yield  the values of six parameters that determine a \lcdm universe. Observations of the PLANCK collaboration provide these values (together with the corresponding 1-sigma spreads); this is the ``universe according to PLANCK" (within the 68\% confidence level, characterized by the $1\sigma$ contours). 

Once these parameters are determined, one can calculate additional observable quantities \emph{assuming} that model and, by carrying out measurements, one can subject the model to consistency tests. For example, the lensing amplitude $A_L$ is set to unity in this construction. One can let this parameter vary and test if this value is consistent with observations. Another type of test is provided by the (odd parity) B-modes. In any one model, one can calculate the correlation function $C^{\BB}_\ell$. As we discuss in Section \ref{s5}, several observational missions will soon measure this correlation function with accuracy that may be sufficient to distinguish one model from another \cite{Matsumura:2013aja,core,Hanany:2019lle}. Similarly, the reionization depth $\tau$ will be measured by missions that are unrelated to the CMB \cite{Fialkov:2016zne}. They will constrain $\tau$, providing us with independent checks on the current \lcdm model.

\subsection{Anomalies}
\label{s2.2}

\begin{figure}
\begin{center}
\vskip-0.2cm
\includegraphics[width=4.2in,height=2.4in]{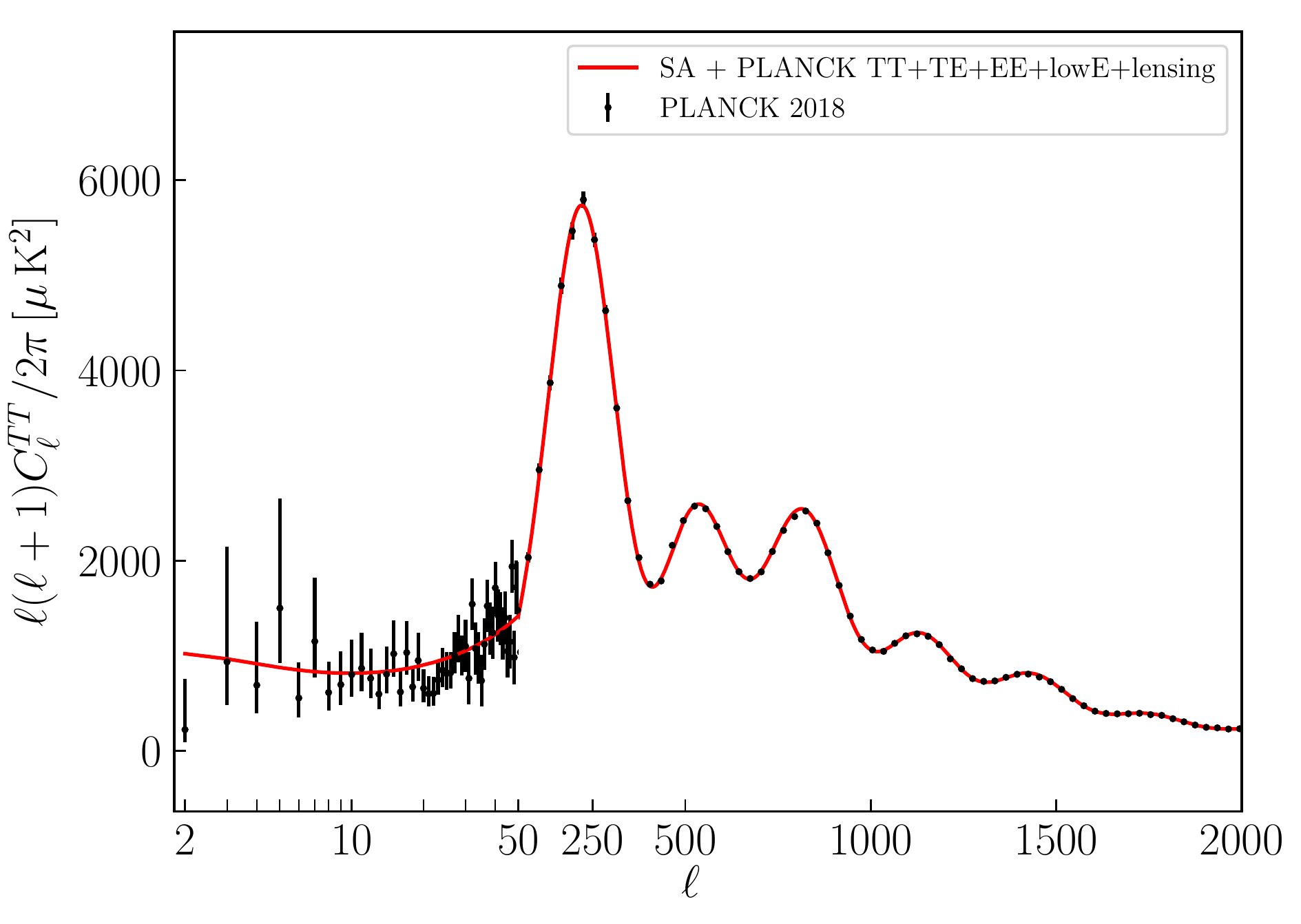}
\caption{\footnotesize{\emph{The TT-power spectrum.} The (red) continuous curve is the theoretical prediction from standard ansatz while the (black) dots with error bars represent the measurements of the PLANCK team reported in 2018. There is excellent agreement between theory and observations for $\ell > 50$ but the observed power is suppressed relative to the theoretical prediction for $\ell \lesssim 30$. As usual, the horizontal axis uses a logarithmic scale for $\ell \lesssim 50$ but a linear scale for $\ell \gtrsim 50$.
}}
\label{CTT-SA}
\end{center}
\end{figure}

\bfig
 \ig[width=0.50\textwidth]{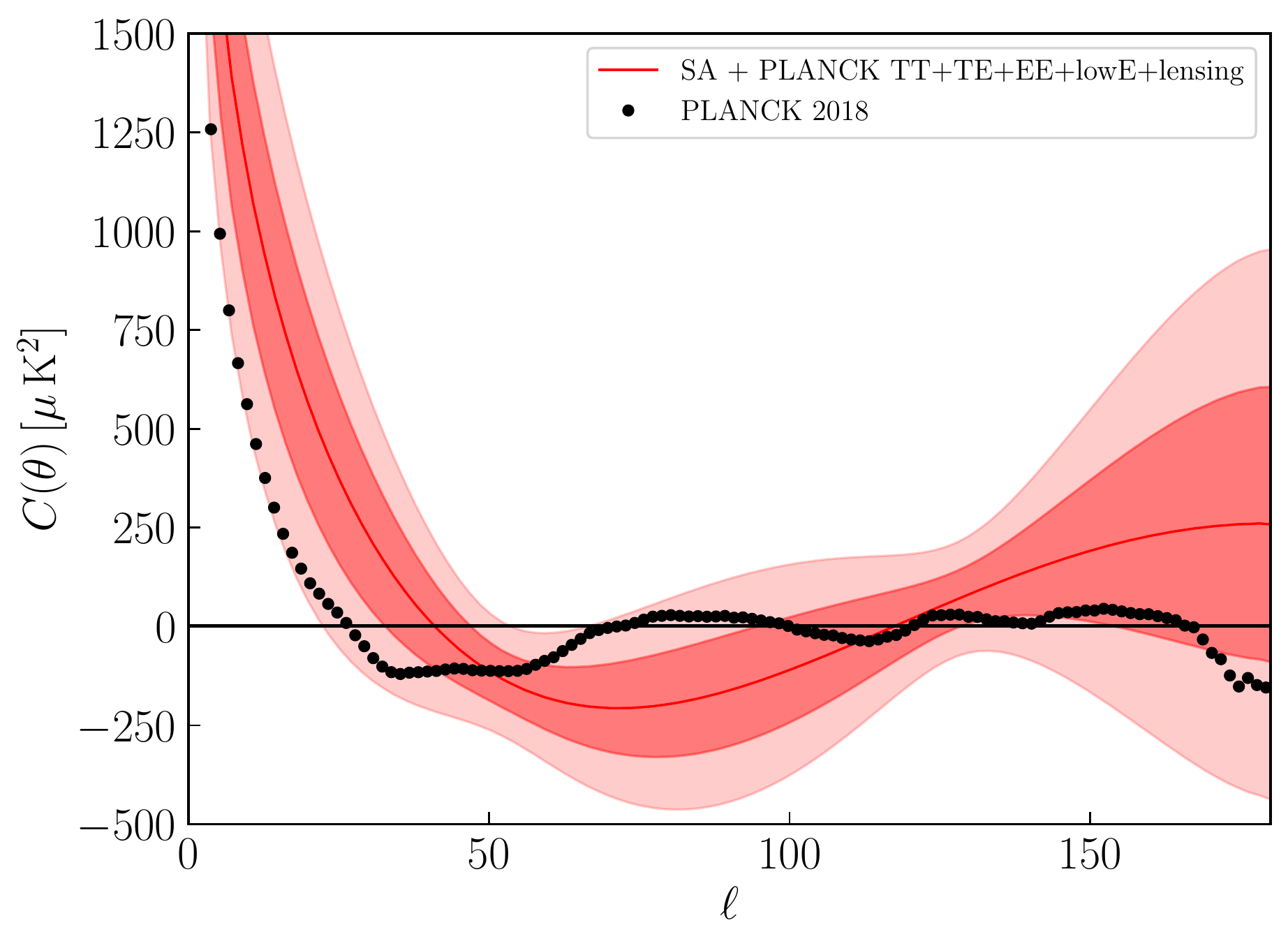}
 \hskip1cm
 \ig[width=0.38\textwidth]{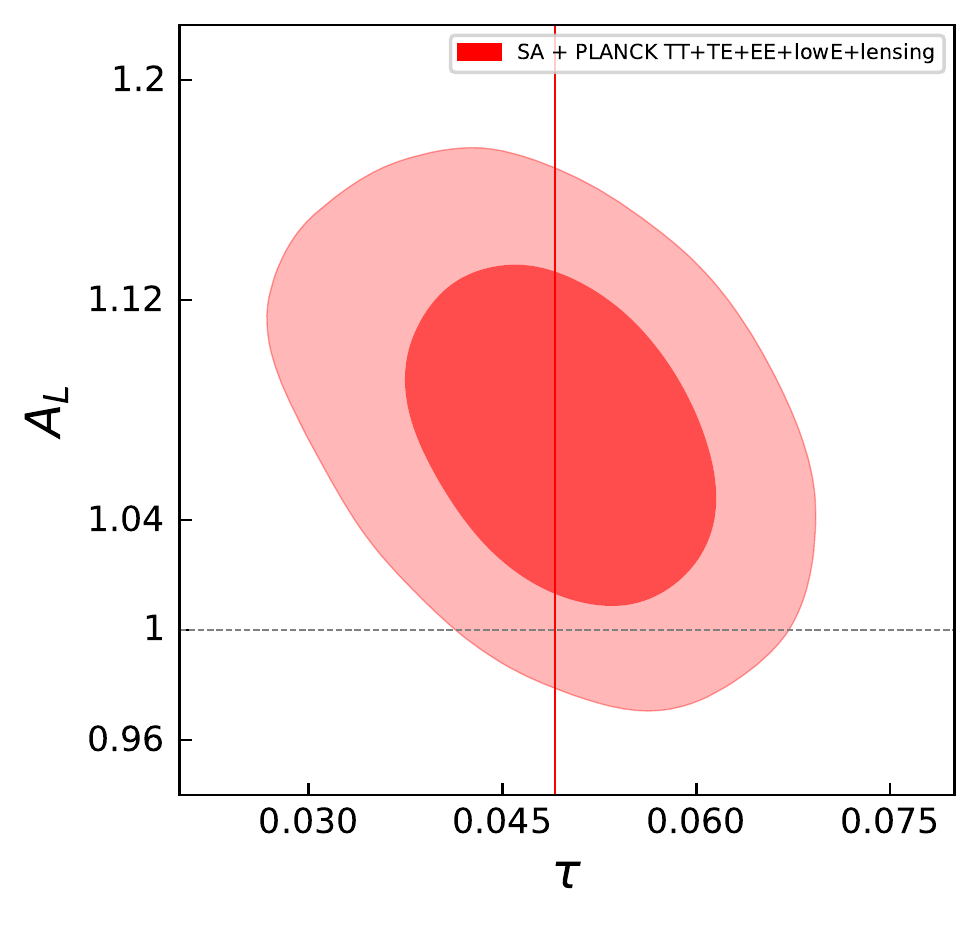}
\caption{\footnotesize{\emph{Left Panel: Large scale power anomaly as measured by $C(\theta)$.} The (red) continuous curve is the theoretical prediction from \SA 
with 68\% and 95\% confidence level contours arising from cosmic variance, while the (black) dots represent the 2018 PLANCK team measurements. The {\it SA} prediction for $S_{1/2}$ is more than 35 times the observed value. \emph{Right Panel: Lensing amplitude $A_L$ versus optical depth $\tau$.} $A_L =1$ lies outside the $1\,\sigma$ contour, signaling the tension between theory and observations.}}
\label{anomalies}
\efig

As Fig. \ref{CTT-SA} shows, the TT power spectrum is in excellent agreement with the theoretical predictions using \SA at small angular scales ($\ell >50$). This is especially noteworthy because the instrumental errors are truly minuscule in this range. 

However, for $\ell \lesssim 30$ the observed power is lower than the theoretical prediction. This power suppression was evident already in the WMAP data, and is reenforced by the PLANCK findings. Over the years it has been argued \cite{wmap,sarkaretal,schwarzetal,planck7} that this anomaly is brought to forefront if one carries out the comparison using the quantity $S_{1/2} := \int_{-1}^{1/2}\left[C(\theta)\right]^2 {\rmd}(\cos\theta)$ that features the physical space TT correlation function $C(\theta)$ in place of the spherical harmonic coefficients $C_\ell$. Qualitatively, large angular scales correspond to small $\ell$ in the spherical harmonic decomposition. However, for any given $\theta_o$ the value $C(\theta_o)$ receives contributions from \emph{all} $\ell$. Therefore
$S_{1/2}$ is a more \emph{direct} measure of the cumulative power for $\theta \geq 60^\circ$ than the $C_\ell$'s for low $\ell$s. Indeed, as the left panel in Fig \ref{anomalies} vividly shows, the observed $C(\theta)$ (the black, dotted curve) is very close to zero for $\theta > 60^\circ$, in contrast to the theoretical (solid, red) curve. The observed value of $S_{1/2}$ is $\sim 1209$, while the theoretical prediction from \SA is $42,496.5$, some 35 times larger. Since the extent of this discrepancy is not immediate from Fig. \ref{CTT-SA}, one may wonder why the power spectrum are not reported using $C^{\TT}(\theta)$ in place of $C_\ell^{\TT}$. The reason is that the $C_{\ell}^{\TT}$ for distinct $\ell$s are (almost) uncorrelated and can therefore be treated as independent observables, while $C^{\TT}(\theta)$ for distinct $\theta$ have massive cross-covariance, whence the statistical significance of power suppression is only $2\!-\!3\sigma$ in spite of the large deviation seen in the left panel of Fig. \ref{anomalies}. Also, because of these correlations, to obtain the 1 and 2$\sigma$ contours in this plot 
one has to take into account a large covariance matrix which in turn requires a detailed understanding of the instruments and the masking procedure near $\theta=180^\circ$ used in the data analysis to remove the contamination coming from the galactic plane. 

The second anomaly is associated with the lensing amplitude $A_L$ depicted in the right panel of Fig. \ref{anomalies}. As it propagates from the last scattering surface at $z\approx 1100$ to us, the CMB is lensed due to inhomogeneities. The lensing potential is nearly Gaussian because there are many lenses along the line of sight. As explained in Section \ref{s2.1}, the six parameter \lcdm universe is determined using best fits to \emph{all four} power spectra. Once this is done, one can compare each observed power spectrum, one by one, with the theoretically predicted power spectrum for that specific \lcdm universe.
Just as this comparison revealed an anomalous suppression of power in $C_{\ell}^{\TT}$ for $\ell <30$, one finds an anomaly also in the lensing potential power spectrum $C_{\ell}^{\phi\phi}$: Relative to the prediction of the best-fit \lcdm model, there is power enhancement in the range $8 \le \ell \le 400$ used by the PLANCK collaboration to report the baseline cosmological results \cite{planck6}. (In this range, the reconstruction procedure is robust and the impact of systematics is reduced). As a consistency check on the 6-parameter \lcdm model, one  introduces a 7th phenomenological parameter $A_{\L}$ --the lensing amplitude, normalized so that $A_{\L} =1$ in the 6 parameter \lcdm model-- and allows it to vary. Varying $A_{\L}$ can be considered as a conservative way of marginalizing over the systematics of the PLANCK data. Departure of $A_{\L}$ from unity signals a tension with predictions based on the standard 6-parameter \lcdm. One finds that $A_{\L}$ is higher than $1$ at $\sim 1.9\sigma$ level \cite{planck6}. This is the lensing amplitude anomaly and its occurrence has been interpreted as a hint of new physics \cite{DivalentinoBridle}. The right panel of Fig. \ref{anomalies} illustrates this tension. Here $A_{\L}$ is plotted against the optical depth $\tau$ because this plot will be useful when we compare the results from LQC with those from the SA in section \ref{s4}: Of the six parameters, only $\tau$ receives significant corrections from LQC. $\tau$ is singled out even within \SA by the fact that the relative error (as measured by the ratio of the standard deviation to the mean value) in  $\tau$ is $\sim 13\%$ while that in the other five \lcdm parameters are less than $1\%$. In the plot, the tension is manifested in the fact that the line $A_{\L} =1$ lies outside the $1\sigma$ contour. Attempts to alleviate this tension within the standard paradigm based on general relativity (GR) --e.g.  changing the background geometry by introducing spatial curvature-- are not supported by lensing reconstruction or Baryonic oscillations (BAO) data (since the joint constraint with BAO is consistent with flat universe, with $\Omega_{\rm K} = 0.001 \pm 0.002$).

As noted in the introduction, while the statistical significance of either of these anomalies is low, together the two imply that the observed universe will emerge only once in $\sim 10^6$ realizations of the posterior probability distributions. Therefore, as the PLANCK collaboration suggested both in its 2015 and 2018 data releases, alleviation of this tension is of considerable interest especially if the mechanism is rooted in physics beyond GR \cite{planck2015xvi,planck1}.

\section{Loop Quantum Cosmology}
\label{s3}

This section is addressed primarily to the cosmology community. In \ref{s3.1} we briefly recall how quantum geometry effects underlying LQG  lead to a resolution of the big bang singularity, replacing it with a big bounce. Since physical quantities do not diverge anywhere, one can extend the standard inflationary scenario all the way to the bounce. In \ref{s3.2}, we discuss the pre-inflationary dynamics of cosmological perturbations, specifically the propagation of quantum fields representing these perturbations on the \emph{quantum} background geometry provided by LQC. In \ref{s3.3}, we explain why, contrary to one's initial expectations, this pre-inflationary dynamics can leave observable signatures at large scales in the CMB. We will use this framework in section \ref{s4} to extract the LQC corrections to the primordial power spectrum. As mentioned in section \ref{s1}, together with observations, these corrections imply that we live in a somewhat different \lcdm universe in which the two anomalies are naturally alleviated.

\subsection{The big bounce of LQC}
\label{s3.1}

Investigations of the early universe are often carried out assuming that spacetime is well approximated by a spatially flat FLRW background metric of GR, together with first order cosmological perturbations that are described by quantum fields. Consider the inflationary paradigm and, for brevity, let us refer to the time when the pivot mode $k=k_\star$ exits the Hubble horizon simply as `the onset of inflation'. At this onset, while spacetime curvature is huge by astrophysical standards --some $10^{65}$ times that at the horizon of a solar mass black hole-- it is only $\sim 10^{-12}$ times the Planck scale. Therefore, at the level of accuracy of current interest, it is safe to ignore the quantum gravity effects even at the onset of inflation. Since spacetime geometry is well approximated by a (perturbed) de Sitter metric at this time, one assumes that the quantum fields representing cosmological perturbations are in the Bunch-Davies (BD) vacuum that is selected by the isometries of the de Sitter metric and evolves the perturbations to the future (as curvature decreases further). 

However, conceptually it is rather ad-hoc to begin, so to say, `in the middle' of evolution. If we go further back in the past, curvature attains the Planck scale, and then diverges at the big bang. During this pre-inflationary epoch, spacetime geometry is \emph{not at all} well-approximated by the de Sitter geometry. Why, then, can we assume the state to be the BD at the onset of inflation? Should we not start in the deep Planck regime and check whether the state is in fact in the BD vacuum at this onset? This would require quantum cosmology, where the Friedmann, Lema\^{i}tre, Robertson, Walker (FLRW) solution of Einstein's equations, characterized by the scale factor $a(t)$ and a matter field $\phi(t)$, is replaced by a quantum state $\Psi(a,\phi)$ subject to an appropriate quantum version of Einstein-matter field equations. Note that reference to the proper time $t$ has disappeared --quantum dynamics is relational, \`a la Leibnitz:  for example, one can use the matter field $\phi$ as an internal clock, and describe how the scale factor evolves with respect to it. Quantum fields representing cosmological perturbations are now to propagate on a \emph{quantum} FLRW geometry $\Psi(a,\phi)$ which assigns probability amplitudes to various metrics, rather than on a single FLRW spacetime.

While this general viewpoint is common to all quantum cosmologies, LQC has two key features that distinguish it from the older Wheeler-DeWitt  (WDW) theory, often called quantum geometrodynamics. First, as explained below, the mathematical framework of LQC descends from the well-developed kinematics of LQG, using a symmetry reduction tailored to homogeneity and isotropy. In the WDW theory one is yet to develop rigorous kinematics for full quantum geometrodynamics; because issues related to the presence of an infinite number of degrees of freedom are generally ignored, the underlying mathematical framework has remained formal. In quantum cosmology, then, one introduces structures like the WDW equation without guidance from a more complete framework. This leads to the second key difference. The LQC quantum Einstein's equation is qualitatively different from the WDW equation, in that it mirrors features of the \emph{quantum} Riemannian geometry of full LQG. As a direct result, strong cosmological singularities --and in particular the big bang-- are naturally resolved in LQG \cite{ps,asrev,30years:IAPS}.

We will now explain these differences in some detail. As is common in quantum field theories, in full LQG one begins with the Heisenberg algebra $\mathcal{A}$ of basic (`canonically conjugate') observables, called the holonomy-flux algebra \cite{aaci,alrev, 30years:KG}. We then have a highly non-trivial result that ensures that $\mathcal{A}$ admits a unique representation by operators of a Hilbert space $\H$ that respects the `background independence' of `diffeomorphism covariance' 
of the theory \cite{lost,cf}. This representation underlies the rigorous kinematical framework of LQG.  In particular, one finds that geometrical observables are well-defined self-adjoint operators with \emph{discrete} eigenvalues. Of particular interest is the \emph{area gap} --the first non-zero eigenvalue $\Delta$ of the area operator. It is a fundamental microscopic parameter of the theory that then governs important macroscopic phenomena in LQC that lead, e.g., to finite upper bounds for curvature.%
\footnote{This is because the curvature operator is defined by considering `Aharanov-Bohm fluxes' across small surfaces $S$ and then shrinking the surface till it has the minimum area $\Delta$.}
In LQC, one first reduces the holonomy-flux algebra $\mathcal{A}$ used in full LQG  to a smaller symmetry reduced algebra $\mathcal{A}_{\rm red}$. Again there is a uniqueness theorem that guarantees that $\mathcal{A}_{\rm red}$ admits a unique representation on a Hilbert space $\Hlqc$ that respects the action of the (residual) diffeomorphism group on $\mathcal{A}_{\rm red}$ \cite{aamc,eht}. This representation is \emph{qualitatively different} (i.e. unitarily inequivalent) from the Schr\"odinger representation used in the WDW theory. In particular, the \emph{differential} operator representing the gravitational part in the WDW equation is not even defined on $\Hlqc$; it is naturally replaced by a certain \emph{difference} operator that explicitly involves the area gap $\Delta$ \cite{apslett,aps,asrev}. One can now start with a quantum state $\Psi(a,\phi)$ that is peaked on the classical dynamical trajectory at a suitably late time when curvature is low, and evolve it \emph{back in time} towards the big bang using either the WDW equation or the LQC evolution equation. Interestingly the wave function continues to remain sharply peaked in both cases. In the WDW theory it follows the classical trajectory all the way into the singularity, while in LQC it ceases to follow the classical trajectory once the curvature is about $\sim 10^{-4}$ times the Planck curvature. Then the quantum geometry corrections dominate and the wave function $\Psi(\phi, a)$ bounces when the curvature and matter density attain their upper bounds. In this \emph{backward evolution}, curvature starts decreasing after the bounce and the universe expands. Once the curvature falls below $\sim 10^{-4}$ times the Planck curvature, the wave function again follows a classical trajectory which is now expanding in the past direction. (For details, see \cite{aps,acs,asrev,30years:IAPS}.)

Thus, key differences between LQC and the WDW theory arise from the fact that the WDW theory has no knowledge of the quantum nature of Riemannian geometry that LQC inherits from LQG. Indeed, there is a precise sense in which the LQC evolution equation reduces to the WDW differential equation in the limit in which the area gap goes to zero \cite{acs}. The upper bound of curvature in LQC is given by ${\rm curv}_{\rm max} =[3\, (24\pi^2)/(2\,{\Delta}^{3})] \,\lp^4 \, \simeq 62 \lp^{-2} $, where, in the last step, we have used the numerical value $\mathring\Delta \simeq 5.17 \lp^2$ of the area gap. In any LQC solution $\Psi_{\tiny{\rm lqc}}(a,\phi)$, the curvature attains its maximum value at the bounce and this value is extremely well approximated by ${\rm curv}_{\rm max}$ if the state is sharply peaked. Note that the upper bound diverges as $\Delta \to 0$, in line with the finding that curvature grows unboundedly as one evolves the WDW state $\Psi_{\tiny{\rm wdw}}(a,\phi)$ to the past. By contrast, in LQC, while the quantum geometry effects are negligible away from the Planck regime, they become dominant in the Planck regime, creating an effective repulsive force of quantum origin that causes the universe to bounce. 

It is interesting that this force rises and falls extremely rapidly, making the agreement with GR excellent outside the Planck regime. However, it has a very non-trivial global effect, in that physics does not stop at the big bang as in GR. Rather, there is an expanding  FLRW universe to the future of the bounce and a contracting FLRW universe to the past, joined by a `quantum bridge'. These qualitatively new features arise without having to introduce matter that violates any of the standard energy conditions, and without having to introduce new boundary conditions, such as the Hartle-Hawking `no-boundary proposal'; they are consequences just of the quantum corrected Einstein's equations. Thus, the existence of the bounce and the upper bound on curvature and matter density can be directly traced back to quintessential features of quantum geometry. These considerations have been extended beyond the spatially flat FLRW models to include spatial curvature, non-zero cosmological constant, anisotropies (see, e.g., \cite{asrev,30years:IAPS} and references therein) as well as the simplest inhomogeneities captured by the Gowdy models in GR, and also to the Brans-Dicke theory (see, e.g., \cite{gowdy,bdtheory}). Taken together, these results bring out the robustness of the LQC bounce.

Since the area gap plays an important role in the LQC dynamics, before concluding this subsection, we will make a small detour to explain how its numerical value $\mathring\Delta \simeq 5.17 \lp^2$ is arrived at. Recall, first, that in QCD there is a quantization ambiguity --parametrized by an angle $\theta$-- because of the freedom in adding a topological term to the action. One encounters a similar quantization ambiguity in LQG (again associated with the freedom to add a term to the action that does not affect equations of motion), encoded in the so-called Barbero-Immirzi parameter, $\gamma >0$, which trickles down to the expressions of observables on $\H$, such as the area operator $\hat{A}_S$. The eigenvalues of $\hat{A}_S$ are discrete in all $\gamma$-sectors. But their numerical values are proportional to $\gamma$ and vary from one $\gamma$ sector to another. Observables also have a $\theta$ dependence in QCD and the value of $\theta$ that Nature has selected is determined experimentally. In LQG, a direct measurement of eigenvalues of geometric operators would determine $\gamma$. But of course such a measurement is far beyond the current technological limits. However one can use thought experiments. Specifically, in LQG the number of microstates of a black hole horizon grows exponentially with the area, whence one knows that the entropy is proportional to the horizon area \cite{abck,abk}. But the proportionality factor depends on the value of $\gamma$. Therefore if one requires that the leading term in the statistical mechanical entropy of a spherical black hole should be given by the Bekenstein-Hawking formula $S= A/4\lp^2$, one determines $\gamma$ and thus the LQG sector Nature prefers. In this sector the explicit value of the the area gap yields $\mathring\Delta \simeq 5.17 \lp^2$ \cite{mdjl,km,30years:FBAP,perez-review} (and the leading term in the entropy of more general black holes --not necessarily spherical-- agrees with the Bekenstein-Hawking formula). This is the value used in LQC calculations.%
\footnote{There are two closely related but technically different ways of characterizing the quantum states of an isolated horizon representing a black hole in equilibrium \cite{30years:FBAP}. They lead to slightly different values of the Barbero-Immirzi parameter ($0.237$ and $0.274$) and hence of the area gap ($5.17 \lp^2$ and $5.98 \lp^2$). Because the values are very close, our results are not sensitive to these differences. See section \ref{s4.3}.}

This concludes our broad-brush overview of how quantum geometry considerations lead to a natural resolution of the big bang singularity in LQC. The resolution has been analyzed in detail in a large number of LQC papers, using Hamiltonian, cosmological-spinfoam and `consistent histories' frameworks (see, e.g., \cite{aps,ach,dcps,asrev,30years:IAPS}). 

\emph{Remark:} Recently some concerns have been expressed about the simplicity of the LQC description of the early universe, and on whether ``general physics principles of effective field theory and covariance" have been appropriately incorporated \cite{mb}. Many of the specific technical points were already addressed, e.g., in \cite{cs,kp,asrev} and in the Appendix of \cite{aa-grg}. In addition, we would like to clarify possible confusion on the following points. First, although `effective equations' are often used in LQC, conceptually they are on a very different footing from those used in effective field theories: One does not integrate out the UV modes of cosmological perturbations. The term `effective' is used in a different sense in LQC: these equations carry some of the leading-order information contained in sharply peaked quantum FLRW geometries $\Psi(a,\phi)$. As we will see in Section \ref{s3.2}, equations satisfied by the cosmological perturbations are indeed covariant. On the issue of simplicity of the LQC description, we note that in the 1980s it was often assumed that the early universe is irregular at all scales and therefore quite far from being as simple as is currently assumed at the onset of inflation. Yet now observations support the premise that the early universe is exceedingly simple in that it is well modeled by a FLRW spacetime with first order cosmological perturbations \cite{inflation-martin}. Therefore, although a priori one can envisage very complicated quantum geometries, it is far from being clear that they are in fact realized in the Planck regime. Nonetheless, one should keep in mind that, as in other approaches to quantum cosmology, in LQC the starting point is the symmetry reduced, cosmological sector of GR. Difference from the Wheeler-DeWitt theory is that one follows the same systematic procedure in this sector as one does in full LQG. But the much more difficult and fundamental issue of systematically \emph{deriving} LQC from full LQG is still open mainly because dynamics of full LQG itself is still a subject of active investigation. See, e.g., \cite{LQGLQC1,LQGLQC2} as illustrations of the current status. 

\subsection{Cosmological perturbations in the pre-inflationary era of LQC}
\label{s3.2}
 
In inflationary paradigms the Mukhanov-Sasaki scalar modes%
\footnote{In the pre-inflationary epoch, the curvature perturbation $\h\R$ for scalar modes become ill-defined at the turn-around point where $\dot\phi =0$. Therefore, in the LQC literature, one uses the Mukhanov-Sasaki gauge invariant scalar perturbation $\h{Q}$ in the pre-inflationary dynamics and converts the result to $\h\R$  at the end of inflation.}
of cosmological perturbations are represented by quantum fields $\h{\Q}$ that propagate on a background FLRW metric ${g}_{ab}$. The use of a classical background geometry is justified since, as explained above, spacetime curvature is twelve orders of magnitude below the Planck scale even at the onset of inflation. However, to extend the paradigm all the way to the LQC bounce, one has to replace the metric ${g}_{ab}$ of GR with an LQC wave function $\Psi(a,\phi)$ because assumptions underlying quantum field theory (QFT) on curved spacetimes fail in the Planck regime. At first the task seems daunting: How do you evolve quantum fields when you have only a probability distribution $\Psi(a,\phi)$ for various spacetime geometries rather than a single metric ${g}_{ab}$? Fortunately, there is an unexpected simplification \cite{akl,aan1,aan3}: So long as $\Psi(a,\phi)$ is sharply peaked, and the back reaction of the perturbations $\h{\Q}$ on the background quantum geometry $\Psi$ remains negligible, dynamics of quantum fields $\h{\Q}$ on $\Psi$ is extremely well-approximated by that of quantum fields  $\h{\Q}$ propagating on a smooth, quantum corrected FLRW metric ${\t{g}}_{ab}$ which is \emph{constructed in a precise manner} from $\Psi$. As one would expect, coefficients of ${\t{g}}_{ab}$ depend on $\hbar$. In the literature, ${\t{g}}_{ab}$ is often called the \emph{dressed metric}. It is `dressed' by certain quantum fluctuations in $\Psi(a,\phi)$ specified below; it carries the information in the quantum geometry $\Psi(a,\phi)$ that the propagation of cosmological perturbations is sensitive to. 

The construction of the dressed metric ${\t{g}}_{ab}$ can be summarized as follows. Recall first that in the standard inflationary scenario, the Mukhanov-Sasaki quantum field $\h{\Q}$ satisfies a wave equation $(\Box + \U/a^{2}) \h{\Q} =0$  where $\Box$ is the d'Alembertian w.r.t. to the background FLRW metric ${g}_{ab}$ (satisfying the unperturbed, zeroth order Einstein's equations) and $\U$ is constructed from the inflationary potential and the background FLRW solution (see, e.g.,\cite{aan3}). At the classical level, this evolution equation can be derived starting with the full Hamiltonian constraint of GR coupled with the scalar field, and then appropriately truncating it to second order in perturbations \cite{aan2}. In LQC, the background quantum geometry $\Psi(a,\phi)$ satisfies the zeroth-order LQC Hamiltonian constraint. The scalar mode $\h{\Q}$ propagates on this $\Psi(a,\phi)$ and its dynamics is governed by the appropriate second order truncation of the full Hamiltonian constraint. If the state $\Psi(a,\phi)$ is sharply peaked and the back reaction of the perturbation $\h{\Q}$ is negligible, then one has the following result \cite{akl}: Propagation of $\h{\Q}$ on the quantum geometry $\Psi(a,\phi)$ is very well approximated by that of a quantum field $\h{\Q}$ satisfying $({\t\Box} + \t\U/\t{a}^2)\, \h{\Q} = 0$. Here ${\t\Box}$ is the d'Alembertian with respect to  the dressed metric 
\be \label{qcg} {\t{g}}_{ab} {\dd}x^a {\dd}x^b \equiv {\dd}\t{s}^2 =
\tilde{a}^{2} (-{\dd}\t{\eta}^{2}\, + \,  {\dd}{\vec{x}^2} )\, \ee
with
\be \label{qpara} 
\tilde{a}^4 = \f{\langle \hat{{H}}^{-\f{1}{2}}\,
\hat{a}^4(\phi)\, \hat{{H}}^{-\f{1}{2}}\rangle}{\langle
\hat{{H}}^{-1}\rangle}\qquad {\rm and} \qquad
{\dd}\tilde{\eta} = \langle \h{{H}}^{-1/2}\rangle\, (\langle \h{{H}}^{-1/2}\, \h{a}^{4}(\phi)\, \h{{H}}^{-1/2} \rangle)^{1/2}\,\, {\dd}\phi\,\, \ee
and $\t{\U}(\phi)$ is the dressed effective potential
\be \label{qpot} \t{\U}(\phi) = \f{\langle \h{{H}}^{-\f{1}{2}}\,
\h{a}^2(\phi)\, \h{\U}(\phi) \h{a}^2(\phi)\, \h{{H}}^{-\f{1}{2}}
\rangle}{\langle \hat{{H}}^{-\f{1}{2}}\, \hat{a}^4(\phi)\,
\hat{{H}}^{-\f{1}{2}}\rangle}\,\,\, . \ee
All operators and their expectation values refer to the Hilbert space of the background FLRW quantum geometry: the expectation values are taken in the state $\Psi$,\, $\h{{H}}$ is the `free' Hamiltonian in absence of the inflaton potential,\, and $\h{a}(\phi)$ is the (Heisenberg) scale factor operator \cite{akl,aan3}. 

At first, the result seems surprising. But physically it can be understood using a simple analogy with propagation of light in a medium such as water. In the full quantum description, individual photons interact with the molecules of the material. However, the key features of propagation can be extracted simply by computing a few macroscopic parameters such as the refractive index and birefringence that can be extracted from the microstructure of the material. Other details of the quantum state of the medium are not important to study propagation. In this analogy, the cosmological perturbation plays the role of light and quantum geometry, the role of the medium. To determine the propagation of $\h{\Q}$, one needs to extract only $\t{a},\, \t{\eta}$ and $\t\U$ from the quantum state $\Psi(a,\phi)$. The rest of the very rich information of quantum geometry it contains is not directly relevant. (Incidentally, the tensor modes satisfy the wave equation for the \emph{same} dressed metric ${\t{g}}_{ab}$; as in standard inflation, there is no dependence on the potential.)

It is clear from the form of equations (\ref{qpara}) and (\ref{qpot}) that the expressions of the dressed metric and the dressed potential could not have been guessed a priori. They resulted from explicit, detailed calculations \cite{akl,aan3}. The observable predictions reported in section \ref{s4} are obtained by first calculating the dressed metric and the dressed potential starting from the given quantum geometry $\Psi(a,\phi)$ and then evolving the scalar mode using $(\t\Box + \t\U/\t{a}^2)\, \h{Q} = 0$.\\

\emph{Remarks:}\\
1. For clarity, let us spell out the conceptual elements of the procedure used to extract dynamics of cosmological perturbations since there is occasional confusion on this point. The starting point in LQC is the Hamiltonian formulation of GR coupled to the inflaton (albeit in the connection variables used in LQG). One then extracts the sector of full GR that corresponds to the homogeneous isotropic fields (which serve as the background) \emph{together with} first order perturbations. It is this classical theory that is then quantized using LQG techniques \cite{aan2}. Dynamics of quantum perturbations are governed by the Hamiltonian constraint operator of the truncated sector, where both the background geometry and perturbations are treated quantum mechanically. One does not simply assume that perturbations satisfy linearized equations of GR on a bouncing classical metric. That the dynamics of perturbations is well approximated by quantum field satisfying an evolution equation involving $\t{g}_{ab}$ and $\t{\U}$ \emph{is a result} that holds under conditions spelled out above. Note also that the equation is covariant w.r.t. $\t{g}_{ab}$ and $\t{g}_{ab}$ rapidly tends to the classical FLRW metric of GR outside the Planck regime.

2. Initially, analysis of \cite{akl} suggested that the propagation of $\h\Q$ on the quantum geometry $\Psi(a,\phi)$ would be exactly the same as that on the corresponding dressed metric \ref{qcg}  and potential (\ref{qpot}) for \emph{any} $\Psi(a, \phi)$. However, Kaminski later found \cite{wk} that there is a subtle infrared problem (that can be missed in numerical simulations since they have to use an infrared cutoff). Kaminski, Kolanowski and Lewandowski \cite{kkl} then showed that, as a result, the implications of \cite{akl} are not as general; the result would not hold without restrictions on the background quantum geometry $\Psi(a,\phi)$. This situation is qualitatively similar to that, e.g., in quantum electrodynamics which also faces infrared issues in rigorous treatments. 
However, in QED the ensuing difficulties can be avoided by focusing just on those quantities that are `infrared safe'. One can adopt a similar strategy in LQC by introducing suitable infrared safe observables through regularization. Furthermore, for states $\Psi(a,\phi)$ that are sufficiently sharply peaked, the regularization ambiguity is completely negligible. Previous calculations of power spectra in LQC (e.g., \cite{aan1,aan3,sab,agullomorris,agulloassym,aks}), as well as the current investigation, use states that are sufficiently sharply peaked in this sense, whence the use of dressed metric is justified in spite of the infrared difficulties.

\subsection{Primordial spectrum: Why pre-inflationary dynamics matters}
\label{s3.3}

A natural question now is whether the pre-inflationary phase of LQC dynamics described in the last two subsections has any observable consequences. Let us therefore focus on the observable modes $\h\Q_k$. These have co-moving wavenumbers $k$ in the range $\sim \, (0.1 k_{\star},\, 300 k_{\star})$, where $k_\star = 0.002 {\rm Mpc}^{-1}$ is the WMAP pivot scale. The evolution equation for these modes implies that, they `experience' curvature in the background metric ${\t{g}}_{ab}$ only if their \emph{physical} wavelength $\lambda(t) = a(t)/k$ is comparable or larger than the radius of curvature $\rcurv (t) = (6/R)^{\f{1}{2}}$ of ${\t{g}}_{ab}$ corresponding to the scalar curvature $R$ at that time. Let us denote by $t= t_\star$ the time at which the relevant slow roll phase starts; this is our onset of inflation. Therefore, a few e-folds before and after $t_\star$, the observable modes propagate as though they are in flat spacetime and therefore do not get excited by the background geometry. What happens in the distant past?  The left panel of \fref{fig:rcurv} shows the evolution of $\rcurv$ (blue solid curve) and of $\lambda$ of observable modes (the gray shaded band), both in GR. Note that in the pre-inflationary epoch $\rcurv$ is far from being constant whence the spacetime metric is very different from the de Sitter metric. Since the scalar curvature $R$ diverges at the big bang, $\rcurv$ goes to zero. Because the scale factor $a$ of the classical FLRW metric goes to zero, physical wavelengths $\lambda$ also goes to zero at the big bang. However, they do not go to zero as fast as $\rcurv$. Therefore, as one approaches the big bang in the past evolution, \emph{all} observable modes exit the curvature radius, `experience' curvature at sufficiently early times and get excited. These excitations have to be delicately fine-tuned for the state to be in the BD vacuum later on, at $t=t_\star$ i.e. at the onset of inflation. Put differently, the Heisenberg state representing the BD vacuum at the onset of inflation is an unnatural choice from the perspective of the Planck regime because it carries certain delicately choreographed excitations there. Of course, one can argue that the quantum field theory in curved spacetime cannot be extrapolated to the Planck regime. But by itself this argument does not provide a justification for using the BD state at $t=t_\star$ either.

\bfig
 \ig[width=0.45\textwidth]{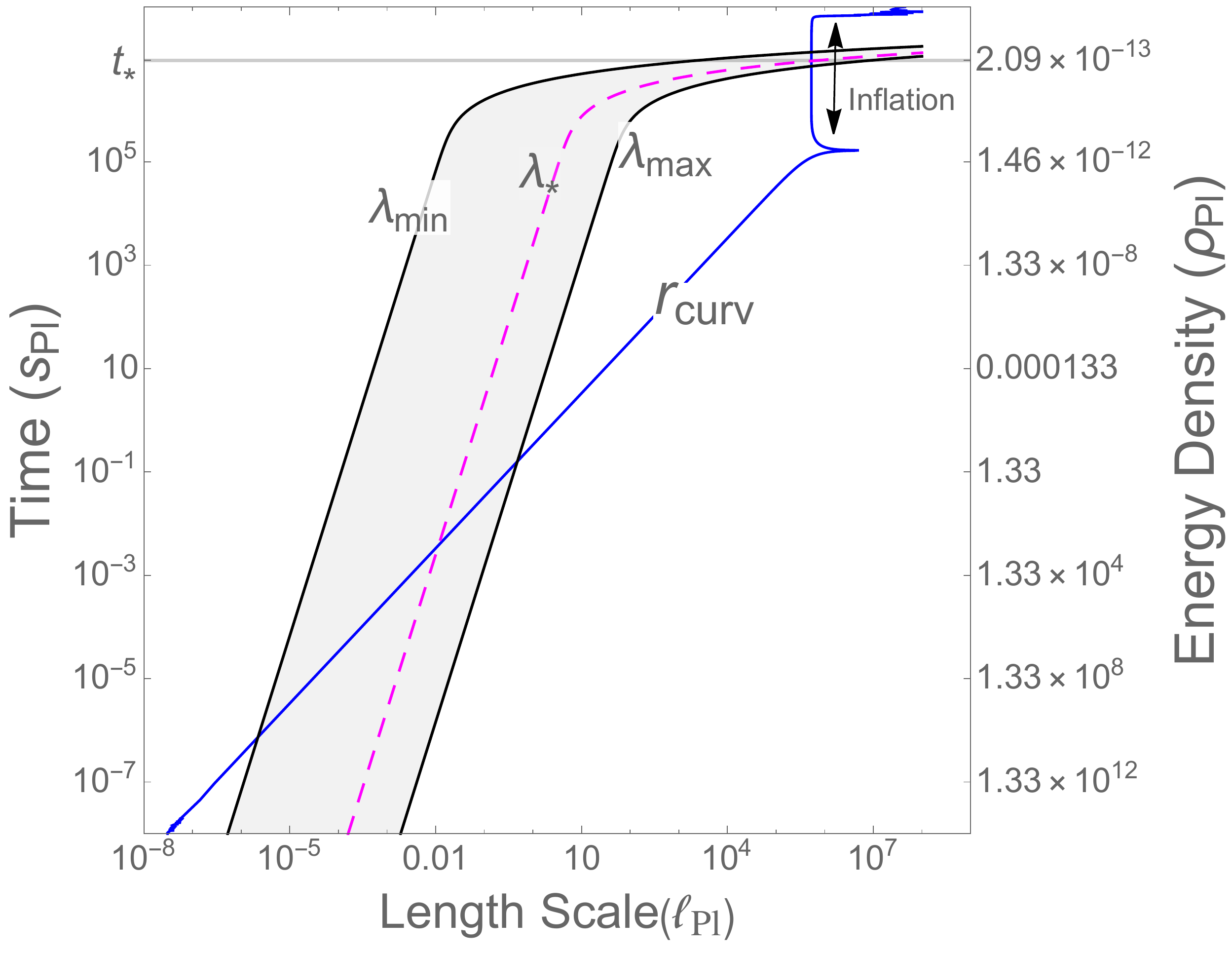}
 \hskip1.2cm
 \ig[width=0.45\textwidth]{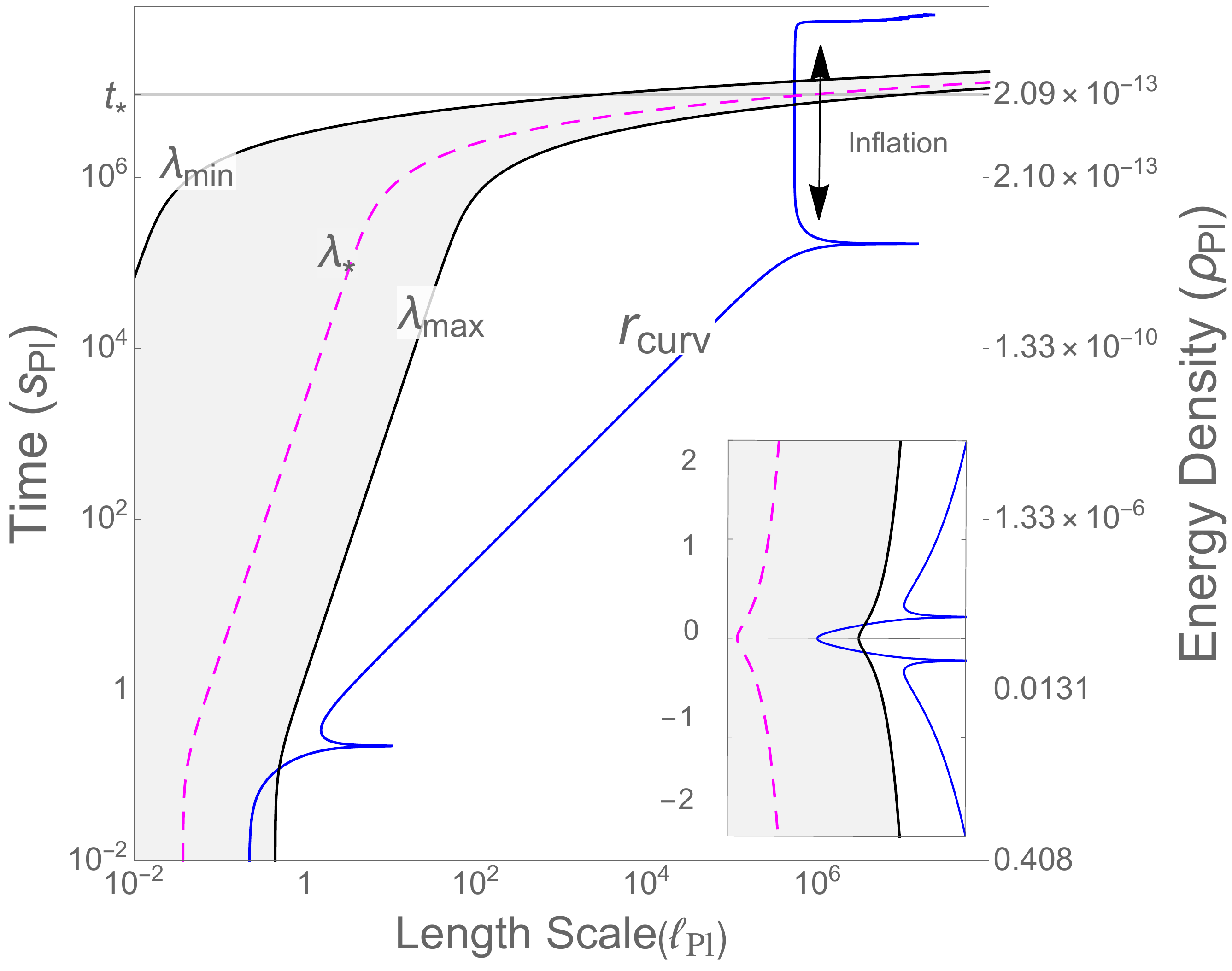}
\caption{\footnotesize{Time dependence of the \emph{physical} wavelengths $\lambda =a/k$ of modes and radius of curvature $\rcurv$ in the pre-inflationary era, using Starobinsky potential. The left vertical axis shows cosmic time $t$ (in Planck seconds) and right vertical axis shows the energy density (also in Planck units). The shaded bands represent the wavelengths of observable modes and the dashed line denotes the WMAP pivot mode with $\lambda_\star =a(t)/k_\star$. The solid (blue) lines represent the evolution of $\rcurv$.
\emph{Left panel: General Relativity.} In the Planck regime near singularity ($t=0$), \emph{all} observable modes exit the curvature radius and are thus excited. 
\emph{Right panel: LQC.} Only the longest wavelength modes in the observable band get excited and fail to be in the BD vacuum at $t=t_\star$. The inset shows dynamics near the bounce.} Plots of the quadratic potential are very similar.} 
\label{fig:rcurv} 
\efig

In LQC, the situation is quite different. Because the scalar curvature $\t{R}$ of the dressed metric $\t{g}$ has a finite upper bound  $\t{R}_{\rm max} \simeq 62 \lp^{-2}$, reached at the bounce, $\rcurv$ reaches its minimum value $\rcurv^{\rm min} \simeq 0.31 \lp$, whence it is only those modes which satisfy $\lambda \gtrsim \rcurv^{\rm min}$ at the bounce that experience curvature in their evolution from the bounce to the onset of inflation. In our approach, (as discussed below) the background quantum geometry $\Psi(a,\phi)$ is such that only the longest wavelength observable modes satisfy this inequality. This feature is shown in the right panel of \fref{fig:rcurv}. Therefore, all but the longest wavelength observable modes propagate from the bounce-time to the onset of inflation as though they are in flat spacetime and hence it is natural that they be in the BD vacuum at $t=t_\star$. It is only the longest wavelength modes that will be excited and hence not in the BD vacuum. But won't these excitations get just washed away during inflation? The answer is in the negative: because of spontaneous emission, the number density of these excitations remains constant \cite{ap,gk}. As a result, as we will see in section \ref{s4}, the primordial spectrum does differ from that based on the {\it SA}, but only at largest angular scales.

Note that there is a deep interplay between the UV and IR in LQC. As we saw in section \ref{s2.1}, in LQC it is the UV modifications of GR in the Planck regime that tame the  big bang singularity and make all physical quantities finite. As a result we have a finite $\t{R}_{\rm max}$ and a non-zero $\t{r}_{\rm curv}^{\rm min}$. It is then natural for all but the longest wavelength observable modes to be in the BD vacuum at the onset of inflation. While the LQC corrections to the background geometry are significant in the UV, their effect on cosmological perturbations is non-negligible only in the IR. This is the point that was highlighted in the abstract and section \ref{s1}. 

Finally, to obtain specific predictions, one needs a quantum state of geometry $\Psi(a,\phi)$ and a quantum state $\psi(\Q,\phi)$ of scalar modes. At this point, different approaches within LQC make different choices (see, e.g., \cite{aan3,madrid,bcgmrev,aaab,agullomorris,agulloassym,menaetal,abs,sab,bjmm,aos1,aos2,aks1,aks}). In this paper, we use the procedure introduced in \cite{ag2,ag3}. Strategy is to select these states by introducing some 
trial principles that relate properties of quantum geometry in the Planck regime with the late time geometry (which can be taken to be that given by general relativity to an excellent degree of approximation). One can then work out the observable consequences. If any prediction is ruled out by observations, one would return to the drawing board and seek alternate principles that would lead to viable states. If predictions are confirmed by observations, one would build confidence in the general direction and attempt to put the principles on a firmer and more satisfactory footing. The currently used principles are somewhat analogous to the Bohr model of the hydrogen atom in the early days of quantum mechanics. While in retrospect it is naive in some fundamental respects, nonetheless the Bohr model was useful because it captured some essential features of the final, correct description of the hydrogen model. Given any inflationary potential, the first principle constrains $\Psi(a,\phi)$, determining the number of e-folds in the pre-inflationary dynamics, i.e., between the LQC bounce and the onset of inflation \cite{ag3}. This restriction enables one to determine the \emph{physical} wavelength $\lambda$ of observable modes at the bounce-time, and hence the angular scale on which the LQC corrections to the power spectrum are significant (which turned out to correspond to $\ell \lesssim 30$ for the potentials considered.) The band of these modes is depicted by the (gray) shaded region in the right panel of \fref{fig:rcurv}, together with its relation to the curvature radius $\rcurv$, depicted by the solid (blue) line. Finally, the principle that determines the quantum state $\psi(\Q, \phi)$ of scalar modes \cite{ag3} involves a quantum generalization of Penrose's Weyl curvature hypothesis \cite{rp-weyl} in the Planck regime near the bounce, which physically corresponds to requiring that the state should be `as isotropic and homogeneous in the Planck regime, as the Heisenberg uncertainty principle allows'.%
\footnote{This condition provides a small ball in the space of all quasi-free states and the desired state $\psi(\Q, \phi)$ is the one in this ball that is `maximally classical' at the end of inflation in a specific, well-defined sense \cite{ag2}.}
While the first principle sets the scale at which the LQC primordial spectrum ceases to be nearly scale invariant, and thus differs from that given by the {\it SA}, the second led to the conclusion that there is power suppression with respect to \SA rather than power enhancement. Detailed calculations are needed to obtain the precise degree of suppression.

\section{Results} 
\label{s4}
In this section we present the main results of this paper, obtained using the LQC summarized in section \ref{s3}. In terms of more commonly used wavenumbers, the new length scale $r_{\rm cuv}^{\rm min} \sim 0.31 \lp$ introduced by LQC provides a new physical scale $k_{\rm LQC} \simeq 3.21 \lp^{-1}$, and the primordial spectrum differs from \SA for $k_{\rm phy} \lesssim k_{\rm LQC}$ at the bounce. In section \ref{s4.1} we first discuss these LQC corrections and then their effect on the observed $C_\ell^{XY}$ correlations (where $XY$ refers to TT, TE, EE and $\phi\phi$). We also present predictions for the value of the optical depth $\tau$ and for the BB power spectrum that could potentially be tested using future observations. In \ref{s4.2}, we show that the LQC predictions for the $C_\ell^{XY}$'s lead to resolution of power suppression and lensing amplitude anomalies. In \ref{s4.3}, we will show that the interplay between LQC and observations is a 2-way bridge, in that the CMB observations can also be used to constrain the value of the area gap $\Delta$, the most important of fundamental microscopic parameters of LQG.

\subsection{Power spectra}
\label{s4.1}
In our LQC model, the physical principles used to select the background quantum geometry imply that the corresponding \lcdm universe has undergone approximately $141$ e-folds of expansion since the quantum bounce until today \cite{ag3}. Therefore, the characteristic LQC scale $\klqc$ corresponds to the co-moving wavenumber $k_o \simeq 3.6 \times 10^{-4}~\mpc$ which sets the scale below which LQC corrections to the primordial scalar power spectrum become important. In particular, the calculations show that for scales $k\lesssim 10k_o$ the power is suppressed whereas for scales $k\gg k_o$ the power spectrum is essentially scale invariant as in the {\it SA}. This behavior is captured in following modification to the {\it SA} for the primordial power spectrum:
\be \label{lqc}
\mathcal{P}^{\rm LQC}_{\mathcal{R}}(k) =f(k)~A_{s}\, 
\left(\f{k}{k_{\star}}\right)^{n_{s}-1}\!\! =\, f(k)\, \mathcal{P}^{\rm SA}_{\mathcal{R}}(k)\, ,
\ee
where $f(k)$ is the correction factor (which is equivalent to the ratio of the power spectra in LQC to that in the {\it SA}). 

\bfig 
\ig[width=2.7in,height=1.7in]{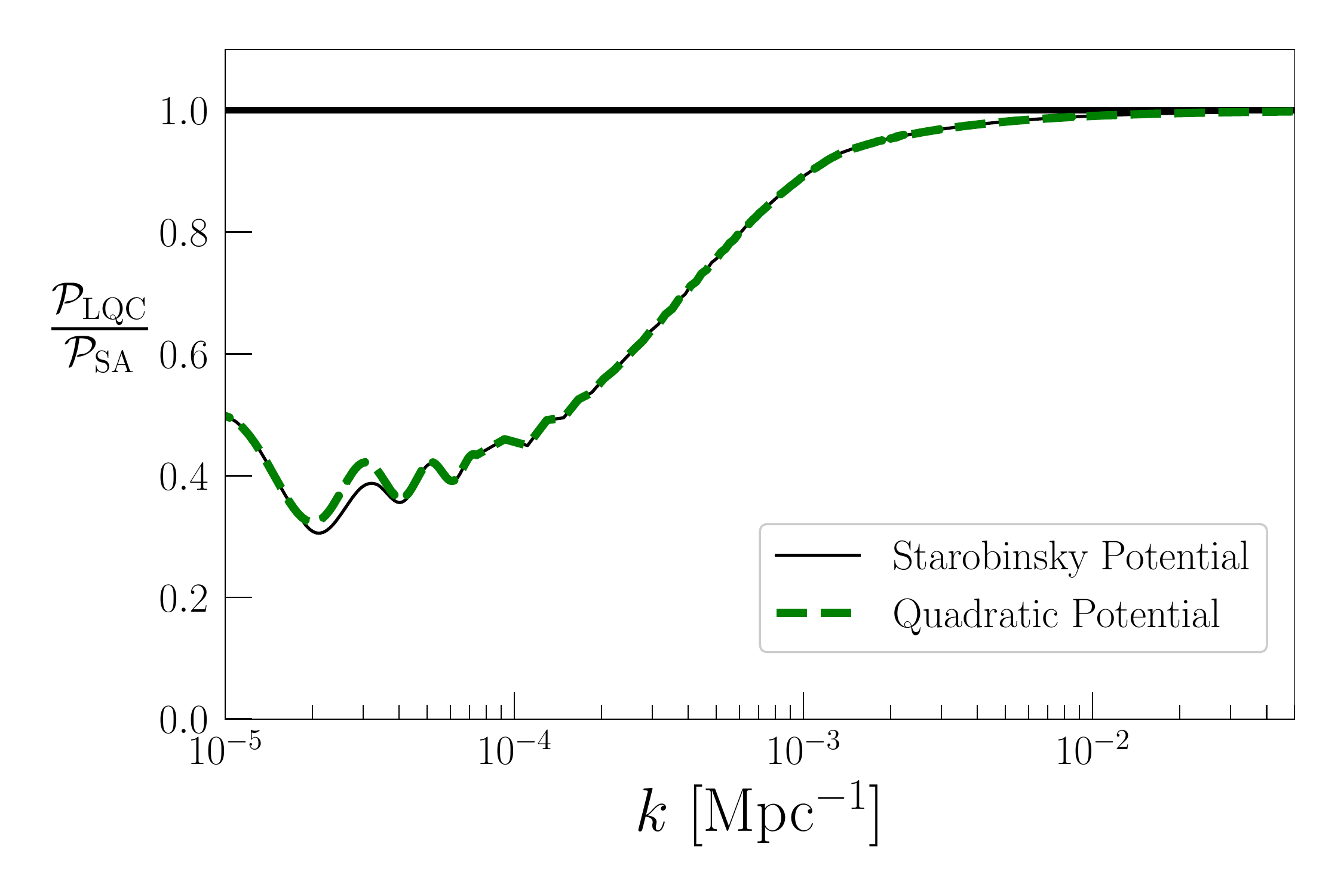}
\hskip0.8cm 
\ig[width=2.9in,height=1.6in,]{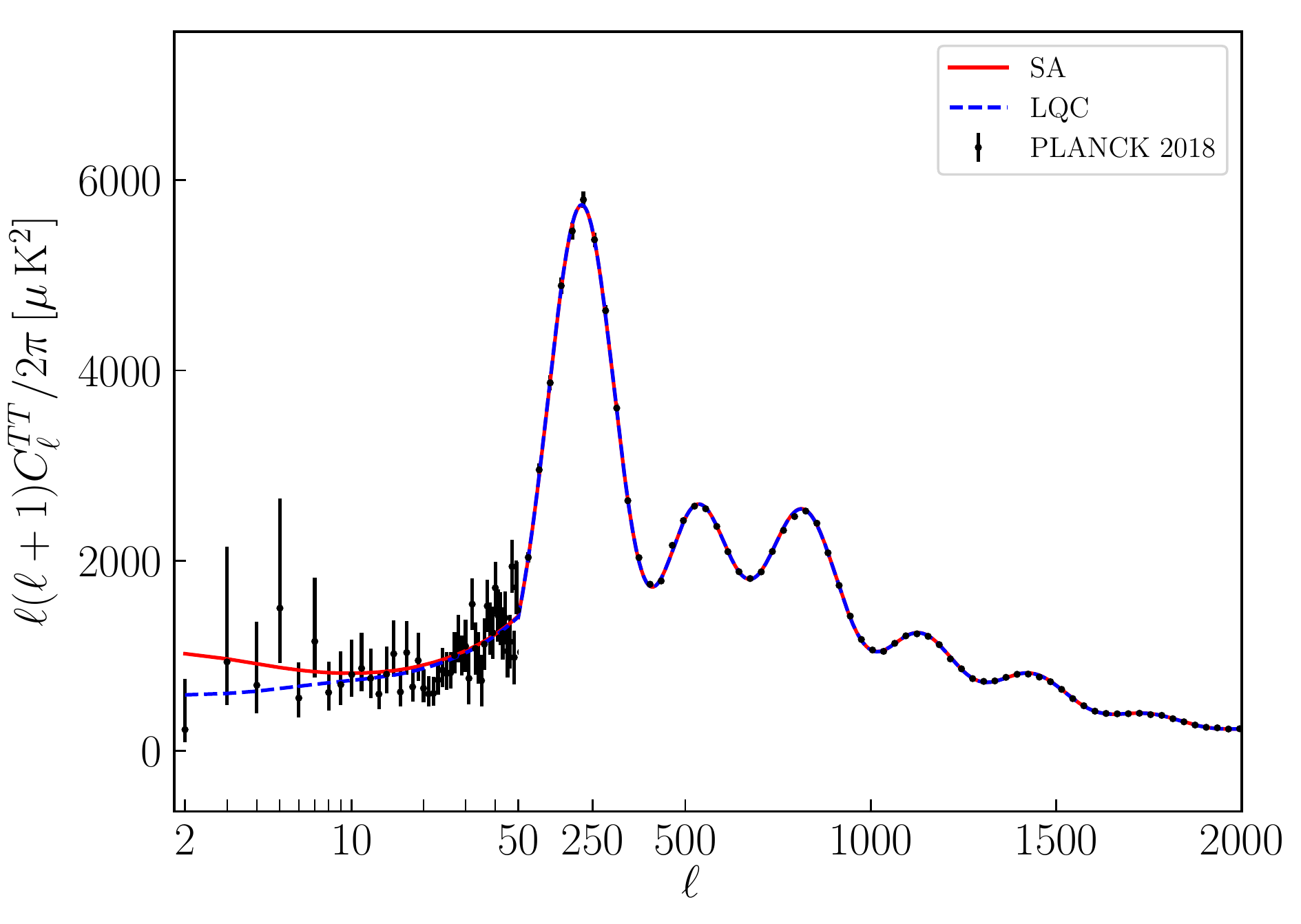}
 \caption{\emph{Left Panel: The primordial power spectrum of LQC}: The LQC suppression factor $f(k)$ is less than $1$ for $k\lesssim 10 \kz \simeq 3.6\times 10^{-3}{\rm Mpc}^{-1}$ for both the Starobinsky and quadratic potentials. At short wavelengths (large $k$), or small angular scales, the LQC power spectrum is indistinguishable from the SA   i.e.\ $f(k) = 1$ for $k \gg k_o$. \emph{Right Panel: TT power spectra.} The 2018 PLANCK spectrum (black dots with error bars),  the LQC (dashed (blue) line) and the standard ansatz  (SA) predictions(solid (red) line). As is usual, on the x-axis we have used a logarithmic scale for $\ell \le 50$ and linear scale for $\ell >50$.}  
\label{fig:power}
\efig

The left panel of Fig. \ref{fig:power} shows the correction factor $f(k)$ plotted with respect to the co-moving wavenumber. It is evident from the plot that for both the Starobinsky and quadratic potentials, relative to the prediction of the {\it SA}, there is power suppression for long wavelength modes corresponding to $k \lesssim 10 k_o$. The origin of this difference lies in the fact that, during the pre-inflationary era, the physical wavelength of these modes is sufficiently large to `experience' the background curvature in the Planck regime near the bounce, leading to excitations over the standard Bunch-Davis (BD) state at the onset of inflation. On the other hand, the physical wavelengths of modes with $k\gg k_o$ are much smaller than the curvature radius throughout the pre-inflationary phase, including the deep Planck regime near the bounce. Therefore their state is practically the same as the standard BD state at the onset of inflation. Note that even for modes with $k \lesssim 10k_{o}$, LQC corrections are significant only near the bounce. During this epoch the energy density of the scalar field is dominated by the kinetic term. Therefore one would expect the inflationary potentials to have negligible effect on the evolution of the background geometry and perturbations in the deep Planck regime, which is when the LQC corrections are imprinted on the modes of scalar perturbations. This expectation is explicitly borne out in the left panel in Fig \ref{fig:power}: the primordial spectra for Starobinsky and quadratic potentials are essentially identical. Analytical considerations of \cite{Copelandetal} suggest that this feature will persist for a large class of inflationary potentials. Finally, note that the LQC primordial spectrum has a turnaround at very large wavelengths  whose origin is related to the sudden spike in $r_{\rm curv}$ near the bounce. Although these modes are not in the observable range for CMB, it is of interest to better 
understand the origin of this growth in power for very small $k$ because, together with large and strongly scale dependent non-Gaussianity (which is expected from the LQC bounce \cite{abs,sab}) it could lead to a coupling between the long wavelength modes and the modes observable in CMB, resulting in a modulation of the primordial power spectrum of the observable modes. Such a non-Gaussian modulation could explain the dipolar modulation of CMB and preference for odd parity that has been observed in the CMB \cite{aks1,aks}. 
\\

Let us now turn to the LQC predictions for \emph{observable} power spectra. For definiteness, in these plots we work with the Starobinsky potential for \SA and LQC theoretical predictions. On the observational side, `PLANCK 2018' refers to the TT+TE+EE+lowE+lensing 2018 data released by the PLANCK collaboration. In order to obtain constraints on the parameters in LQC and SA models, we used the publicly available software package \texttt{COSMOMC} \cite{Lewis:2002ah} which supports the likelihood code used in the original 2018 PLANCK analyses. \texttt{COSMOMC} is based on the Markov-Chain-Monte-Carlo (MCMC) procedure for estimation of parameters based on maximum likelihood analysis. For a given theoretical model with a number of parameters (here the \lcdm model with 6 parameters), one builds a Markov chain from randomly sampled parameter values from a space predefined by priors. Each Markov chain begins with a random sample from the prior-constrained parameter space. Subsequent steps in the chain are then selected via Metropolis-Hasting algorithm: a Monte Carlo sample from the parameter space is proposed and then accepted or rejected based on the likelihood function which quantifies the degree of agreement between the theoretical prediction and the experimental data  (in this paper we work with the same likelihood function as used in PLANCK 2018 papers \cite{planck5}.). This procedure is repeated until a convergence criteria is satisfied. The chain of accepted values are further trimmed and thinned in order to remove dependence on the initial data point and correlation between subsequent steps. The converged Markov chain thus obtained approximates the posterior distribution of the parameters to be estimated. In order to obtain one-dimensional constraints on (or two-dimensional correlation between) the individual parameters, marginalization procedure is used. For details of the MCMC techniques adapted to cosmological settings see \cite{Lewis:2002ah}.

\bfig
\ig[width=0.465\textwidth]{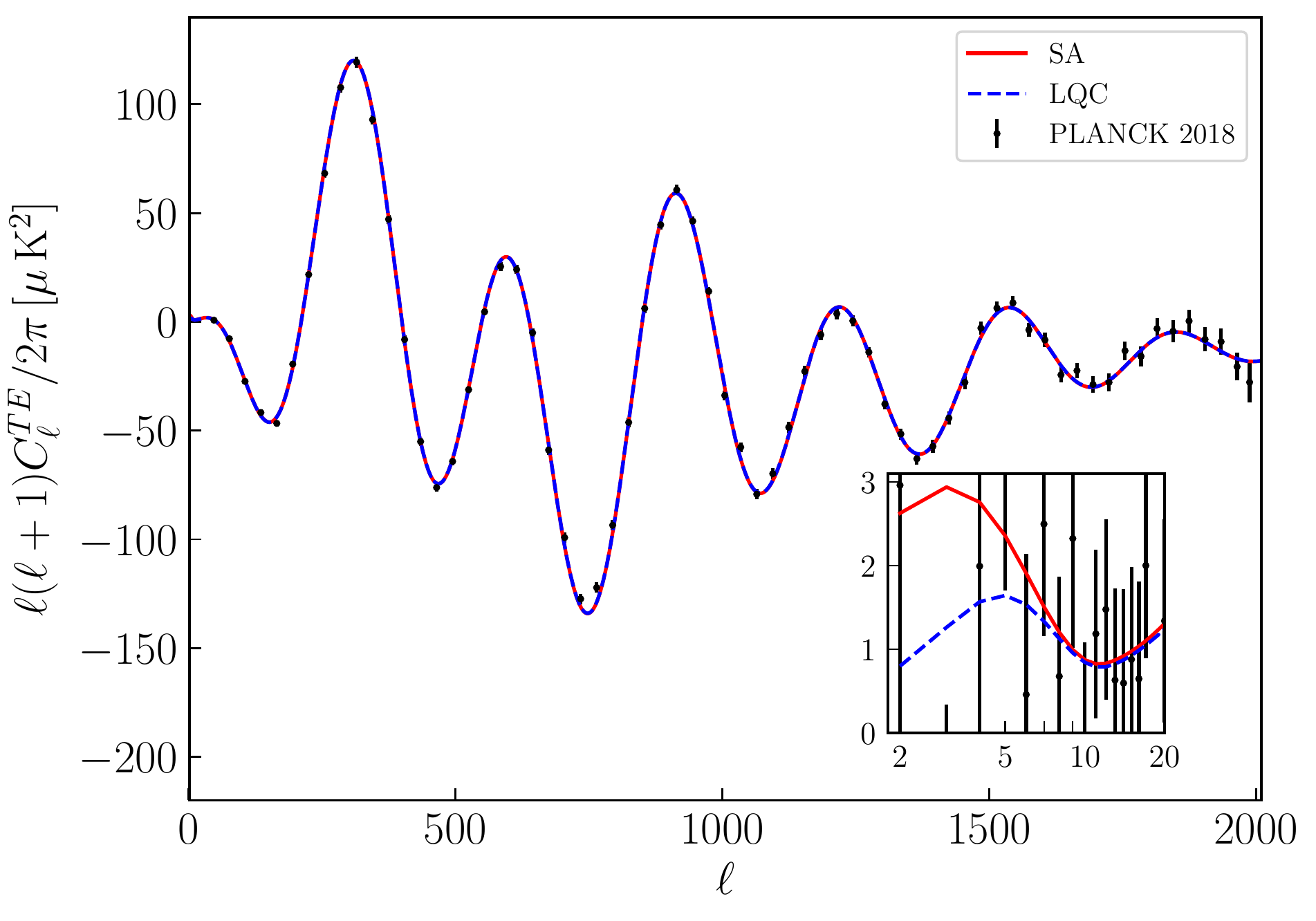}
\hskip1cm
\ig[width=0.45\textwidth]{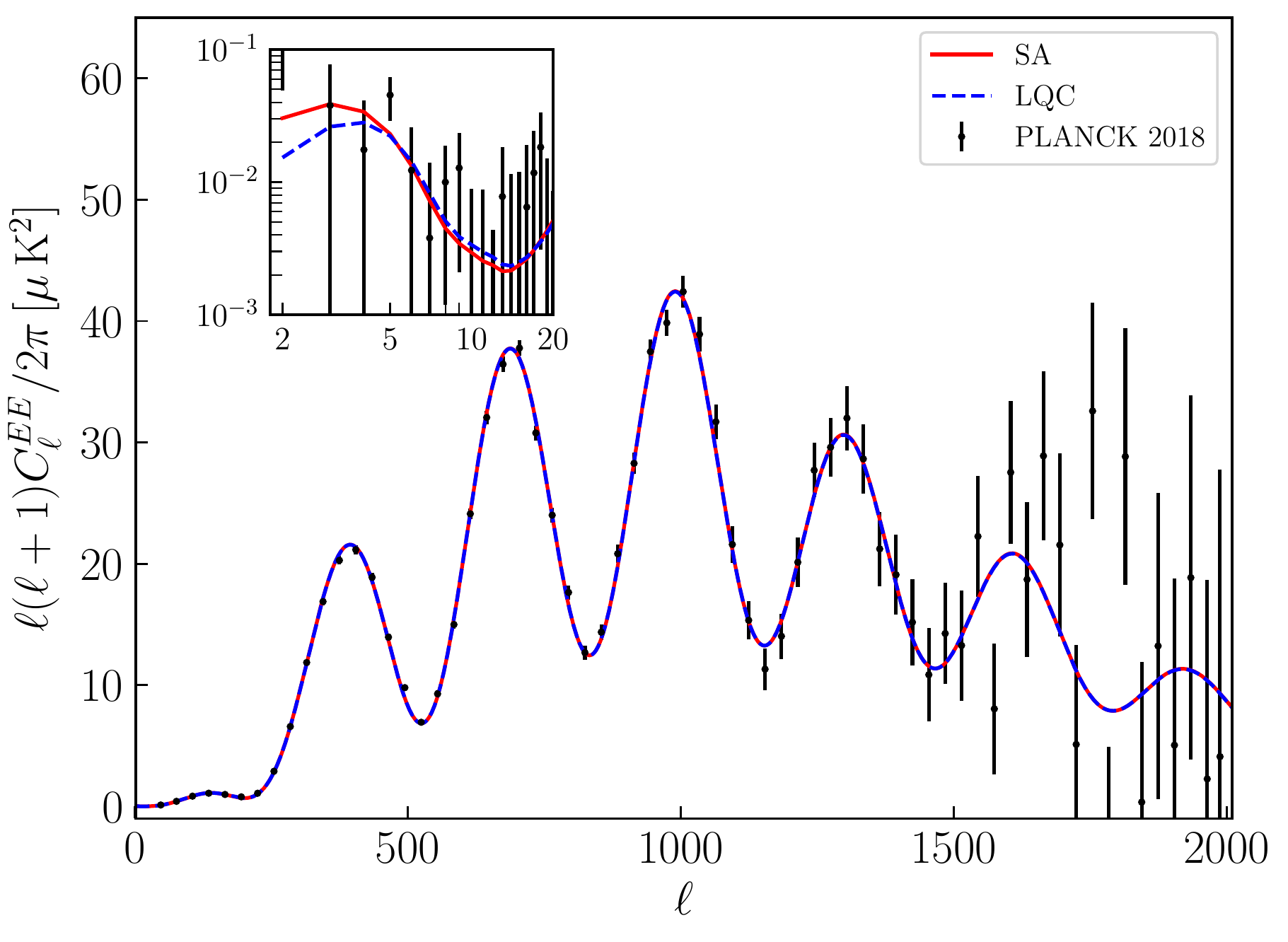}
 \caption{\footnotesize{\emph{TE and EE power spectra.} The 2018 \planck spectrum (black dots with error bars),  the LQC (dashed (blue) line) and the standard ansatz  (SA) predictions(solid (red) line). Values of the 6 \lcdm parameters are fixed to their best fit values shown in table \ref{tab1}.}} 
\label{fig:EE}
\efig

The right panel of Fig. \ref{fig:power} shows the theoretical predictions as well as the observed power spectrum, with its error bars. The plots clearly show that the primordial power suppression of LQC for small $k$ leads to power suppression in the TT spectrum for multipoles $\ell<30$, relative to the predictions from the {\it SA}.
The same primordial power spectrum can also be used to compute the predicted TE and TT correlation spectra. Fig. \ref{fig:EE} shows the TE (left panel) and EE (right panel) spectra as observed by Planck 2018, along with those obtained using the best fit \lcdm models with LQC and the {\it SA}. Note that in the right panel of Fig. \ref{fig:power} and in Fig. \ref{fig:EE} the LQC and \SA plots are shown for \emph{their corresponding} marginalized mean values of the 6 parameters shown in Table \ref{tab1}. By contrast, in \cite{ag3} both the LQC and \SA curves were plotted for the same values of 6 parameters of the \lcdm model, coming from the {\it SA}.  As Table \ref{tab1} shows, the mean value of optical depth $\tau$ is\, $ > 9\%$\, higher in LQC than with the {\it SA} and electric polarization is quite sensitive to $\tau$. Interestingly, this change in the best fit \lcdm universe has the effect of slightly reducing  suppression in the passage from the primordial power spectrum to the observed one. As a result, the LQC power suppression in the TT, TE and EE spectra we now find is somewhat less pronounced than it was in \cite{ag3}.

\begin{table}
\begin{center}
\footnotesize
\begin{tabular}{| c | c | c | }
\hline
{\rm Parameter}   &                         {\rm SA}               &                                               \rm{LQC}
 \\ \hline \hline

$\Omega_b h^2$  &     $0.02238 \pm 0.00014$       &                           $0.02239 \pm 0.00015$ \\ \hline
$\Omega_c h^{2}$    &     $0.1200\pm 0.0012$         &                         $0.1200\pm0.0012$ \\ \hline
$100\theta_{MC}$        &              $1.04091 \pm 0.00031$          &                       $1.04093 \pm 0.00031$ \\  \hline
$\tau$           &              $0.0542\pm 0.0074$             &                     $0.0595\pm0.0079$  \\ \hline
$\ln (10^{10} A_s)$   &    $3.044\pm 0.014 $         &                        $3.054\pm0.015$ \\ \hline
$n_s$               &          $0.9651 \pm 0.0041$               &                      $0.9643\pm0.0042$   \\ \hline
\hline
$S_{1/2}$           &          $ 42496.5$                                                   &               $14308.05$ \\   \hline

$A_L$               &          $1.072\pm 0.041$    
&               $1.049\pm 0.040$\\

\hline
\end{tabular}
\caption{\emph{Comparison between the Standard Ansatz (SA) and LQC.}  The mean values of marginalized probability distributions for the six cosmological parameters, and values of $S_{1/2}$ calculated using $C_{\ell}^{\rm TT}$.}
\label{tab1}
\end{center}
\end{table}

\bfig
\ig[width=0.44\textwidth]{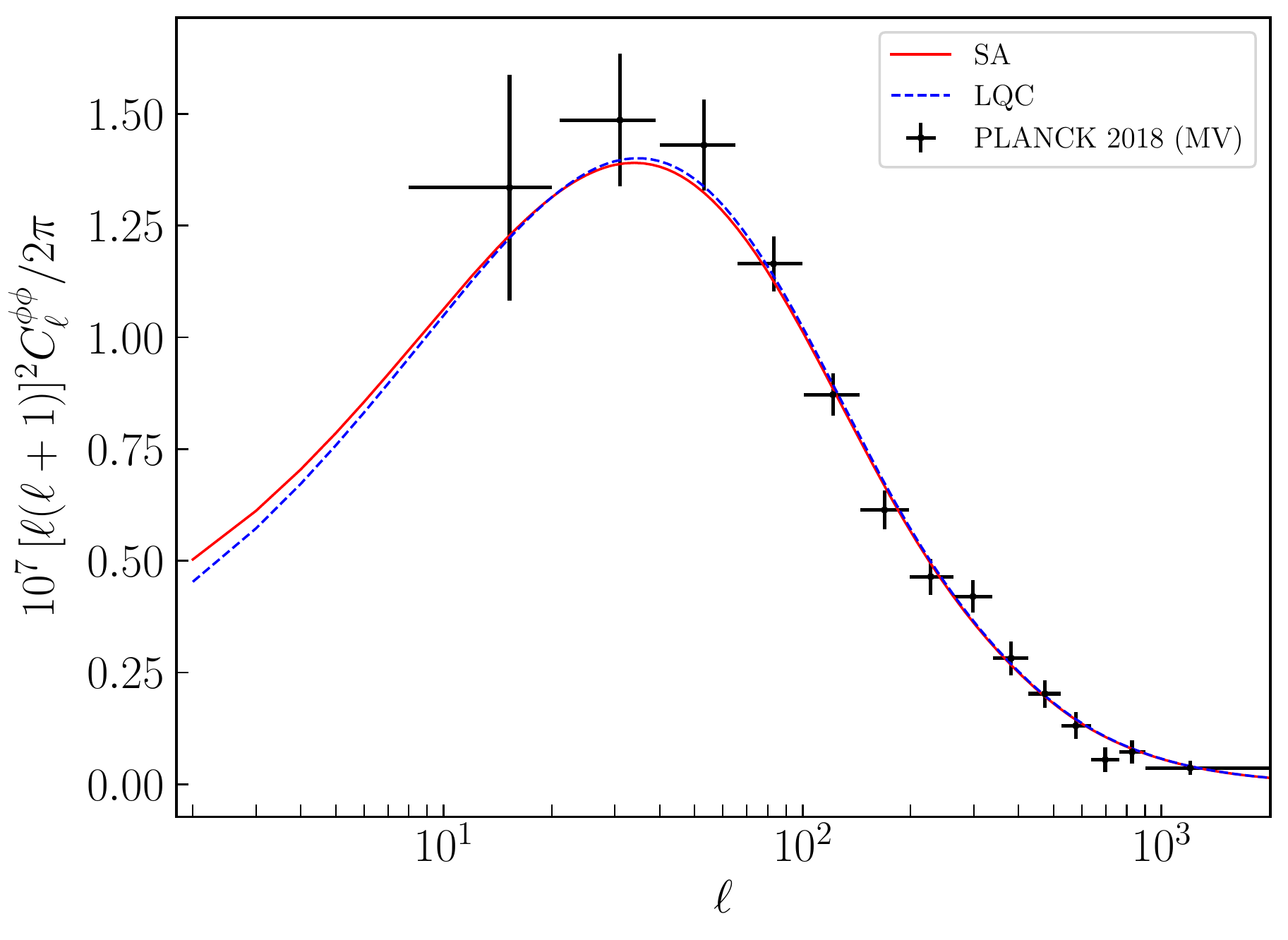}
\hskip0.7cm
\ig[width=0.47\textwidth]{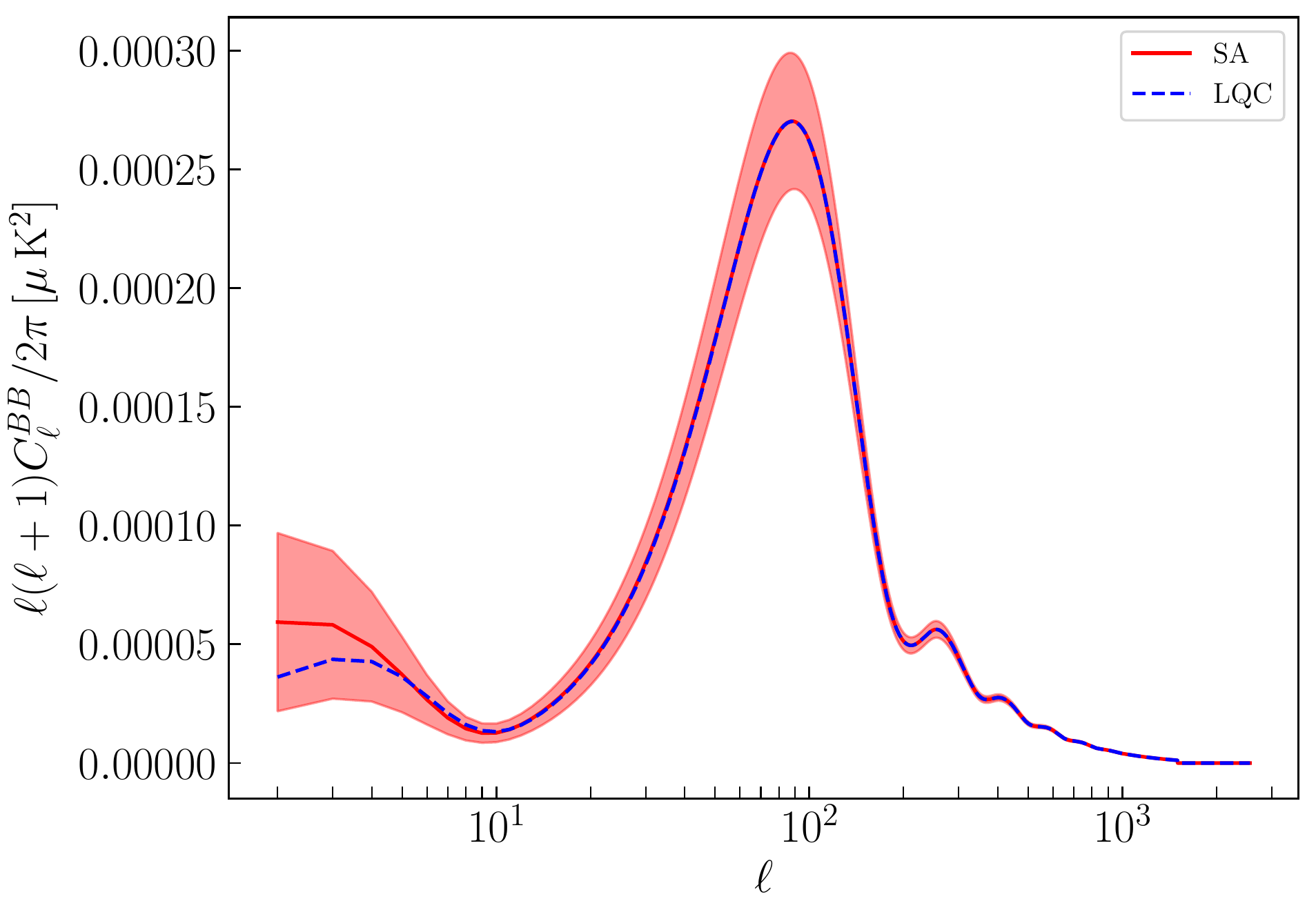}
 \caption{\footnotesize{LQC prediction (dashed (blue) line) and the standard ansatz  (SA) prediction(solid (red) line). \emph{Left Panel: Lensing power spectra.} The 2018 \planck spectrum (where the black crosses show binning and error bars).
 {\emph{Right Panel: The theoretical predictions for the BB power spectra.} There is power suppression in LQC for very low $\ell$. The shaded region shows the 68\% confidence level region. }} }
\label{fig:phiphi}
\efig

Finally, the left panel of Fig.\ \ref{fig:phiphi} shows the lensing correlation spectrum $C_\ell^{\phi\phi}$ reported by the PLANCK collaboration with their 2018 data, along with the LQC and \SA predictions. For this figure we have again  used the mean marginalized values of the 6 parameters shown in Table \ref{tab1}, and the lensing amplitude is fixed to $A_L=1$, in accordance with the base \lcdm model. As with the TT, TE and EE spectra, the lensing spectrum also shows suppression at large angular scales corresponding to $\ell < 30$. However, observations are quite sparse for low $\ell$. Interestingly, for $30<\ell<100$, the LQC prediction for $C_\ell^{\phi\phi}$  is slightly larger than that for SA. (This is in fact inevitable; see section \ref{s5}.) This behavior leads the LQC spectrum to fit slightly better with the observed data in the range $30<\ell<100$ and hints towards resolving the lensing amplitude anomaly in LQC without having to introduce additional modifications to the standard \lcdm model. In the next subsection we will see that this possibility is in fact realized. \\ 

We will conclude this discussion of observable implications of LQC with predictions for future missions. First, as we noted in section \ref{s2.2}, currently the optical depth $\tau$ is the least accurately measured of the 6 \lcdm parameters, with a relative error of $\sim 13\%$. The LQC value is some 9.8\% higher than that in `the universe according to PLANCK'. This prediction will be tested by the future observation of global 21cm evolution at high redshifts that is estimated to reach a percent level accuracy in the measurement of $\tau$ \cite{Fialkov:2016zne}. As for the observable power spectra, to date the PLANCK satellite has provided the best full sky measurement of the CMB anisotropies. However, there is still scope for improvement in the measurement of electric polarization, and the odd-parity magnetic polarization is yet to be detected. Therefore, several space-based mission have been planned to further improve the polarization measurements. In addition to the predictions for the TE and EE power spectra discussed above, our LQC model also makes predictions for the BB power spectrum. The right panel of Fig.\ \ref{fig:phiphi} 
provides the prediction for the unlensed BB power spectrum both from \SA and in LQC. Recall that the tensor-to-scalar ratio $r$ depends on the potential of the inflationary model. But, being a ratio, it is the same in LQC as from SA \cite{aan3}. We have set $r=0.0041$, its value for the Starobinsky potential. As in the case of other four spectra, we observe a relative suppression of power at low multipoles. However, this is also where the reionization bump occurs. Since LQC predicts a larger value of optical depth, the B-B power suppression is lower than what one might have expected from the primordial power suppression (and using the same value of $\tau$ for both SA and LQC). Nonetheless, it may be possible to test this prediction against the data from the future B-mode missions such as LiteBIRD \cite{Matsumura:2013aja}, Cosmic Origins Explorer \cite{core}, ECHO \cite{cmb-bharat} or Probe Inflation and Cosmic Origins (PICO) \cite{Hanany:2019lle} (which should observe the BB spectrum if $r\gtrsim 0.001$).

\subsection{Anomalies alleviated}
\label{s4.2}
\bfig
  \ig[width=0.65\textwidth]{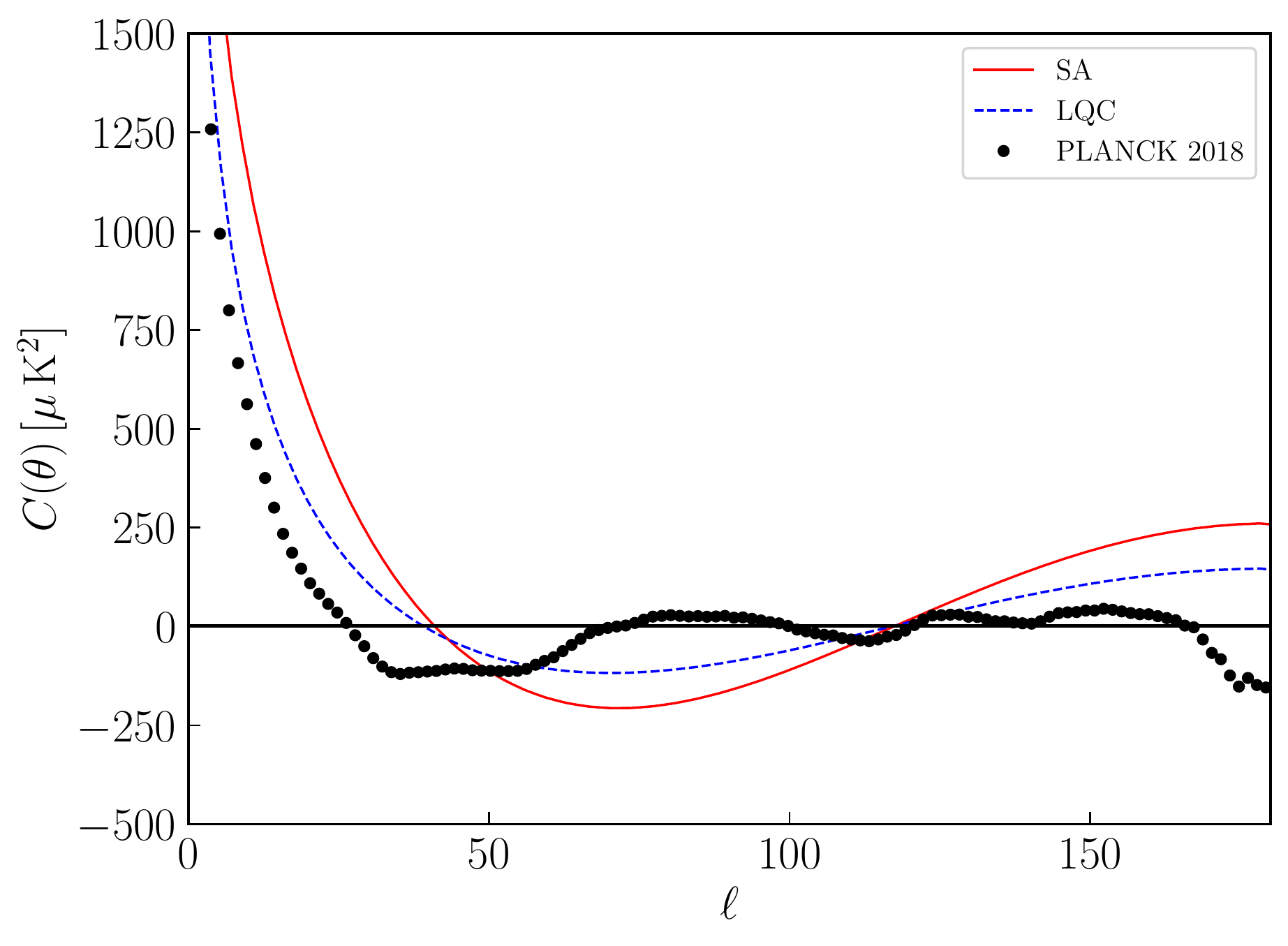}
 \caption{\emph{The temperature-temperature angular power spectrum $C(\theta)$.} The 2018 \planck spectrum (thick black dots),  the LQC  (dashed (blue) line), and the standard ansatz (solid (red) line) predictions. The LQC predictions are closer to the observed Planck 2018 data points.} 
 \label{fig:Ctheta}
\efig

\bfig
 \ig[width=3in]{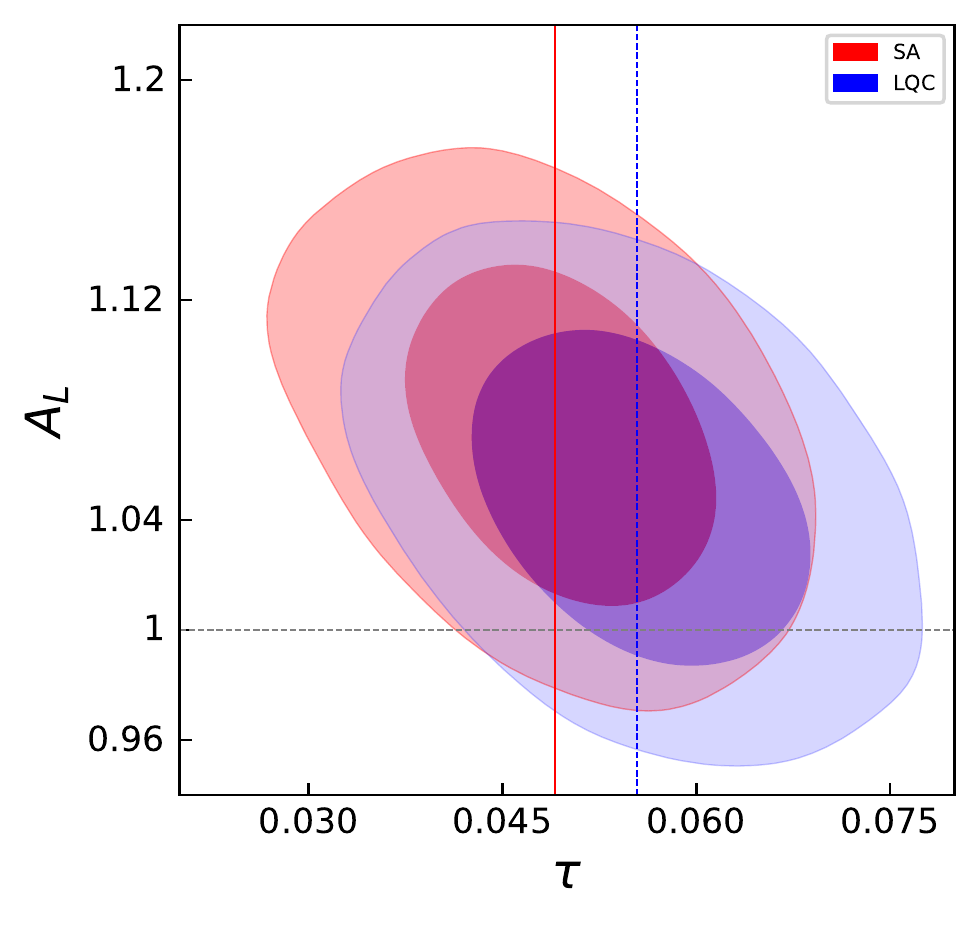}
 \caption{\emph{$1\sigma$ and $2\sigma$ probability distributions in the $\!A_{L}-\tau$ plane.} Predictions of the \SA (in red) and LQC (in blue). Vertical lines denote the mean values of $\tau$ for \SA and LQC. It is evident that $A_L=1$ is outside the $1\sigma$ contour for \SA while is restored within $1\sigma$ for the LQC model.} 
\label{fig:taual}
\efig

As discussed in Sec. \ref{s2.2}, in our approach the pre-inflationary phase of LQC dynamics enables one to address two CMB anomalies. The first, shown in the left panel of Fig.\,\ref{anomalies}, is that the observed angular correlation function $C^{\rm TT}(\theta)$ remains close to zero for angles $\theta > 60^\circ$ in contrast to the predictions of the standard \lcdm model based on the {\it SA}. Let us now examine the prediction of $C^{\rm TT}(\theta)$ from LQC. 
This prediction is plotted as the dashed(blue) curve in Fig.\! \ref{fig:Ctheta}, along with the prediction from \SA shown as the (red) dashed curve and the 
$C^{\rm TT}(\theta)$ observed by the 2018 PLANCK mission shown by (black) dots. One sees by inspection that the LQC predictions are closer to the observed values than \SA predictions. This behavior can be further quantified by comparing the value of $S_{1/2} := \int_{-1}^{1/2}\left[C(\theta)\right]^2 {\rmd}(\cos\theta)$ which, as we saw in section \ref{s2.2}, represents a cumulative total power at large angular scales ($\theta > 60^\circ$): 
\be
S_{1/2}^{\rm Planck} = 1209.18; \qquad 
S_{1/2}^{\rm SA} =  42496.5;\qquad
S_{1/2}^{\rm LQC}  = 14308.05.
\ee
The tension between observations and the theoretical prediction from the \SA is encapsulated by the fact that $S_{1/2}^{\rm SA}$ is almost 35 times larger than $S_{1/2}^{\rm Planck}$. This is the power suppression anomaly. This discrepancy is appreciably reduced in LQC since $S_{1/2}^{\rm LQC}$ is $\sim 1/3$ of $S_{1/2}^{\rm SA}$. As indicted in section \ref{s2.2}, because $C(\theta)$ for different values of $\theta$ are correlated, and because one has to use a masking procedure in the data analysis near $\theta =180^\circ$, the task of providing the $1\sigma$ and $2\sigma$ contours around the LQC plots is challenging, requiring manipulations of a large covariance matrix in the data analysis, as well as a detailed understanding of aspects of the instrument. It would be of considerable interest if the CMB experts could provide these plots starting with the LQC TT-power spectrum. Note also that because power is suppressed for $\ell \lesssim 30$, if LQC results were used in the PLANCK analysis, error bars for low $\ell$ would also be reduced and we would have a sharp measure of the LQC alleviation of this anomaly. 

The second anomaly, shown in the right panel of Fig.\ \ref{anomalies}, is the lensing amplitude anomaly. This anomaly is not directly observed in the $C_\ell^{\rm XX}$ plots or in $C(\theta)$ plots, but arises when one performs a consistency check of the \lcdm model. Instead of fixing the lensing amplitude to $A_L=1$ (as is assumed in standard \lcdm model) one allows it to vary along with the standard 6 parameters, i.e., one now analyzes a 7 parameter model. The anomaly lies in the finding that $A_L=1$ is beyond $1\sigma$ error bar as shown in the right panel of Fig.\ \ref{anomalies}. This led the authors of \cite{silketal} to conclude that there is a possible ``crisis in cosmology" because, to alleviate this problem, one would need to introduce spatial curvature which creates significant departures from the observed power spectra as small angular scales. As shown in Fig. \ref{fig:taual}, however, repeating the analysis with the LQC power spectrum restores $A_L=1$ within $1\sigma$ contour thereby resolving the lensing amplitude anomaly and avoiding the hint of a potential ``crisis". 

To summarize, the observed CMB power spectrum has many non-trivial and interesting features at small angular scales and observational error bars are small in this regime. It is remarkable that these features are correctly predicted by the  \lcdm model based on the SA. The LQC corrections could well have led to discrepancies with these successful predictions of the {\it SA}. That does not happen. Rather, there are departures only \emph{at large angular scales} leading to a suppression of the \emph{primordial} power for small $k$. This difference from the nearly scale invariant standard ansatz (\ref{SA}) leads to the alleviation of two anomalies in the CMB. 

\subsection{From observations to fundamental theory}
\label{s4.3}

\bfig
 \ig[width=0.7\textwidth]{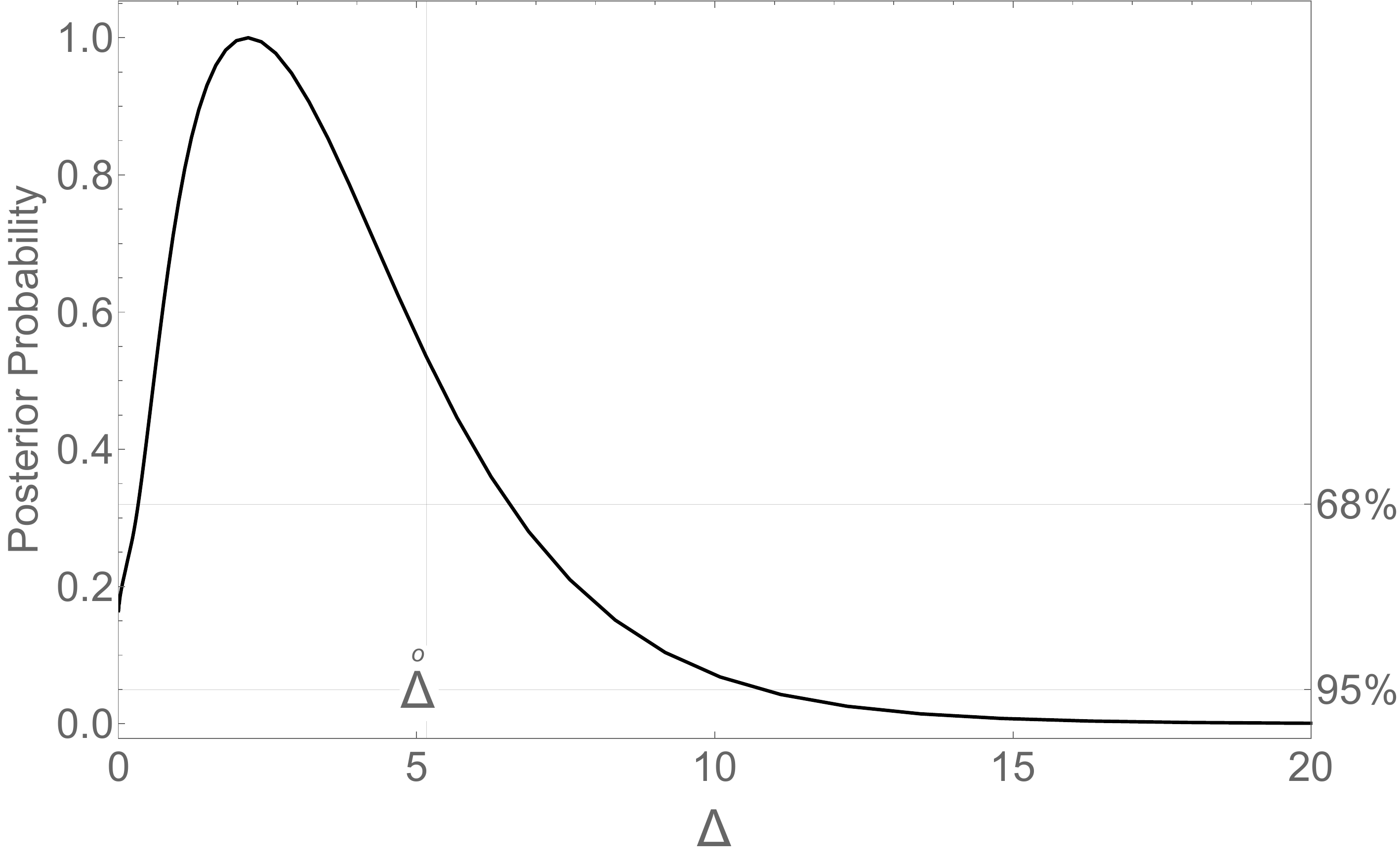}
\caption{\footnotesize{\emph{Marginalized posterior probability distribution for $\Delta$.} The value $\mathring\Delta \simeq 5.17 \lp^2$ selected by the black hole entropy considerations is denoted by the vertical line. It lies within $1\sigma$ of the marginalized mean value $ 3.86\lp^2$ that lies to the right of the peak. Thus, there is an unforeseen internal consistency.}}
\label{fig:RB}
\efig

As we saw in section \ref{s3}, the area gap ${\Delta}$ is the key microscopic parameter that determines values of important new, macroscopic observables such as the matter density and the curvature at the bounce. Its specific value, $\mathring{\Delta} = 5.17~\lp^2$, is determined by the statistical mechanical calculation of the black hole entropy in loop quantum gravity (see, e.g., \cite{abck,abk,30years:FBAP,perez-review}). Results reported in the last two subsections are based on this value. However, with the CMB observations at hand, we can now, so to say, turn the tables and regard $\Delta$ as an additional parameter and \emph{use the CMB observations to constrain its value}.

Thus, let us consider $\Delta$ to be a \emph{free parameter} and obtain a posterior probability distribution for its value by letting it vary along with the 6 parameters of the \lcdm model. Note that this procedure is similar to the self-consistency check that is performed for the lensing amplitude $A_L$. The \lcdm universe determined by the {\it SA} does not pass that test because, when it is allowed to vary, $A_L$ prefers a value that is greater than 1 and the discrepancy is significant in that the value 1 lies outside the $1\sigma$ contour of the best fit value \cite{planck6}. Is there perhaps a similar tension here? More precisely, does the value $\mathring\Delta \simeq 5.17 \lp^2$ obtained from black hole entropy calculations lie within the 68\% confidence contour of the value preferred by the CMB observations? If it does not, LQC would fail the self-consistency test at the $1\sigma$ level. If it does, there would be an unforeseen coherence between detailed conclusions drawn from vary different considerations: the LQG analysis of black hole entropy and the LQC investigation of the very early universe!

Fig.\ \ref{fig:RB} shows the one-dimensional posterior distribution of $\RB$. The 
the best-fit value --the peak in the distribution-- is at $\Delta = 2.18 ~\lp^2$ 
while the marginalized mean value is $3.86\lp^2$.

with the following constraint:

\be  1.26 \lp^2 < \Delta < 6.47\lp^2 \qquad ({\rm at~68\%~confidence~level}).
\ee

Clearly, the value $\mathring{\Delta} \simeq 5.17 \lp^2$ chosen in Sec. \ref{s3.3} and used in this paper, is within 68\% ($1\sigma$) confidence level of the constraint obtained from Planck 2018. This not only indicates a synergy between the fundamental theoretical considerations and observational data, but also provides internal consistency of the LQC model. 

\section{Discussion}
\label{s5} 

To determine the six parameter \lcdm universe we live in, the PLANCK team began with the standard ansatz (\ref{SA}) for the primordial spectrum and used known astrophysics to determine the observable power spectra for various values of the six cosmological parameters, $A_s, n_s, \Omega_b h^2, \Omega_c h^2, 100\theta_{MC}$. By comparing these theoretical predictions with observations, they determined the marginalized mean values of the six parameters (together with the error bars corresponding to 68\% confidence level). This is the \lcdm universe according to PLANCK. This procedure has had tremendous success, especially with the finer  features of the power spectra at small angular scales. However, there are also some anomalies. Since they are only at $\sim 2\!-\!3 \sigma$ level, the statistical significance of any one anomaly is low. However, taken together, two or more anomalies imply that we live in an exceptional realization of the six posterior distributions provided by this procedure. 

One can view these anomalies as potential gates to new physics. Indeed, the PLANCK collaboration has emphasized this possibility in its 2015 \cite{planck2015xvi} as well as 2018 data releases \cite{planck1}. As the second of these papers points out, {\sl ``...if any anomalies have primordial origin, then their large scale nature would suggest an explanation rooted in fundamental physics. Thus it is worth exploring any models that might explain an anomaly (even better, multiple anomalies) naturally, or with very few parameters.''} LQC researchers have followed up on this suggestion. In this paper we presented a concrete realization of this idea. Specifically, one begins with the observation that, in the standard procedure summarized above, the theoretical input, beyond known astrophysics, is the {\it SA}, motivated by the inflationary scenario. It assumes that the primordial spectrum is nearly scale invariant and can be characterized just by two numbers $A_s$ and $n_s$ across \emph{all} wavenumbers $k$. However, in LQC the resolution of the big bang singularity introduces a new scale $k_{\rm LQC}$ and quantum gravity corrections in the pre-inflationary phase of dynamics appear for $k \lesssim k_{\rm LQC}$. For these small wavenumber modes, the LQC primordial spectrum is no longer nearly scale invariant, whence there is departure from the predictions drawn from the {\it SA}. 

Several closely related approaches have been used in LQC to probe the effects of this pre-inflationary dynamics (see, e.g., \cite{aan1,aan2,aan3,madrid,bcgmrev,lcbg,aaab,agullomorris,agulloassym,ag2,ag3,menaetal,abs,bjmm,sab,aks}). Our approach has two main ingredients: (i) the use of sharply peaked quantum states $\Psi(a,\phi)$ for the background \emph{quantum} FLRW geometry, that then enable one to well-approximate the dynamics of cosmological perturbations on the quantum geometry $\Psi(a,\phi)$ by that on a quantum corrected `dressed metric' $\t{g}_{ab}$ \cite{abk,aan3}; and, (ii) the use of certain principles to select $\Psi(a,\phi)$ and the quantum state $\psi(\Q, \phi)$ of the scalar mode of cosmological perturbations for any given inflationary potential \cite{ag3,ag2}. For any given inflationary potential, the principle used to select $\Psi(a,\phi)$ limits the number of e-folds during pre-inflationary dynamics, thereby implying that the modes that receive significant LQC corrections in the primordial spectrum correspond to the large angular scales $\ell \lesssim 30$. The principle used to select $\psi(\Q, \phi)$ implies that in the primordial spectrum there is power reduction (rather than enhancement) in these modes. This then translates to a power suppression for $\ell \lesssim 30$ in the observed power spectra. For modes with $\ell \ll 30$, the LQC power spectra are indistinguishable from those obtained using the {\it SA}. Thus, LQC predictions leave the highly successful predictions of standard inflation at small angular scales unaffected, but modify the predictions at large angular scales. 

Details of the LQC pre-inflationary dynamics reveal some interesting facts. First, the quantum geometry effects on the background FLRW geometry are dominant only in a short interval around the bounce. Second, it is the modes whose physical wavelength $\lambda_{\rm phy}$ is longer than the curvature radius $\rcurv$ during the pre-inflationary evolution that fail to be in the BD vacuum at onset of inflation. In the observable band, only the longest wavelength modes are thus affected and they satisfy $\lambda_{\rm phy} \gtrsim \rcurv$ only for 2-3 e-folds after the bounce. Thus, the background quantum geometry as well as the quantum perturbations Planck receive non-negligible LQC corrections during a \emph{very short duration}. Yet these corrections lead to observable effects in that they alleviate some anomalies. Third, while we did not discuss tensor modes in this paper, their power spectra have the same behavior as that of scalar modes and, given an inflationary potential, the tensor to scalar ratio $r$ does not receive LQC corrections (within accuracies reported here). Finally, there is an unforeseen interplay between the UV and the IR: While it is the \emph{UV modifications} of GR that lead to the singularity resolution and create the new LQC scale $k_{\rm LQC}$, the structure of the evolution equations satisfied by cosmological perturbations is such that it is the \emph{IR modes} with $k \lesssim k_{\rm LQC}$ that are affected during their pre-inflationary evolution. It is this unforeseen cosmic tango between the very small and the very large that is responsible for the alleviation of the two anomalies discussed in this paper.

Given that LQC simultaneously alleviates the power suppression and the lensing amplitude anomalies, it is worth investigating a more general question: Are the two conceptually related? As reported in \cite{agjs}, the answer is in the affirmative. Since this relation seems not to have been noticed before, we will make a small detour to explain it. Let us begin by assuming that there is \emph{some} mechanism --not necessarily originating in LQC--
that provides a  primordial power spectrum of the form 
\be \label{new} \mathcal{P}^{\rm new}_{\mathcal{R}}(k) =f(k)~A_{s}\, 
\left(\f{k}{k_{\star}}\right)^{n_{s}-1} \ee
with $f(k) < 1$ for $k < \kz$ and $f(k) = 1$ for $k > \kz$ for some $\kz$, and compare and contrast the new $\Lambda$CDM universe obtained from this modified ansatz with that given by the SA of Eq.~(\ref{SA}). In the first step, we can restrict our analysis only to smaller angular scales ($k\gg \kz$). Then, the primordial spectrum in both schemes would be the same, whence we  would obtain the same marginalized mean values of the six cosmological parameters. Denote by $\Az_{s}$ the marginalized mean value of the scalar amplitude $A_{s}$ thus obtained.  In the second step, let us consider the \emph{full} range of observable modes, including $k \le \kz$. Now, given that the observations show that the TT power is suppressed at large-scales, i.e., for $k \lesssim \kz$, if one uses the SA+$\Lambda$CDM model the marginalized mean value $A_{s}^{\rm SA}$ using the \emph{entire} $k$ range will be lower than $\Az_{s}$. By contrast if the primordial power spectrum is of the form of Eq.~(\ref{new}), the initial power is already suppressed by $f(k)$. Therefore, $\Az_{s}$ will not have to be lowered as much to obtain the marginalized mean value $A_{s}^{\rm new}$.  Thus, we have 
\be \Az_{s}\, >\, A_{s}^{\rm new}\, >\, A_{s}^{\rm SA}. \ee  
The key point is the last inequality: $A_{s}^{\rm new} > A_{s}^{\rm SA}$. (We spelled out the argument because at first it seems counter-intuitive that power suppression leads to a larger $A_{s}^{\rm new}$. But note that power is suppressed only for low $k$.)  Next, we know that for large $k$, the product $A_{s}e^{-2\tau}$ is fixed  by observations. Hence, it follows that the best fit values of the optical depth in the two scheme must satisfy $\tau^{\rm new} > \tau^{\rm SA}$. Finally, from the very definition of lensing amplitude, the value of  $A_{L}$ is anti-correlated to the value of $A_{s}$. Therefore, it follows that we have the inequality  $A_{L}^{\rm new} < A_{L}^{\rm SA}$. Thus  in any theory that has  primordial spectrum of the form  (\ref{new}),  $A_{s}, \tau$ and $A_{L}$ will have the same \emph{qualitative} behavior as in LQC, and hence the tension with observations would be reduced. What LQC provides is a precise form of the suppression factor $f(k)$ from `first principles', and hence specific quantitative predictions. 
Also, recall from section \ref{s4.1} that in our analysis the LQC $f(k)$ also came with a specific value $k_o \simeq 3.6 \times 10^{-4}~\mpc$ for $\kz$. Other mechanisms could well lead to a very different value. If so, in the observed power spectra suppression would arise at a very different value of $\ell$. Finally, the specific $f(k)$ computed from LQC also leads to other predictions  --e.g., for the BB power spectrum discussed in section \ref{s4.1}-- that need not be shared by other mechanisms. 

In this respect, it would be of interest to compare the LQC predictions with those that result from effective field theories \`a la Ginsburg and Landau, where slow roll inflation is generically preceded by a fast roll phase that leads to a suppression of CMB quadrupole \cite{bddvs}. This will require a calculation of angular power spectrum $C(\theta)$, and of the measure $S_{1/2}$ of power suppression, using the marginalized posterior probabilities of the 6 cosmological parameters in this effective field theory approach, and a reanalysis of the lensing amplitude along the lines of section \ref{s4.2}. Yet other mechanisms have been proposed to account for power suppression at large angular scales in the context of GR (see, e.g. \cite{isw1,isw2,ContaldiFastRoll,ClineFastRoll,JainFastRoll,PedroFastRoll,LelloFastRoll,cai3}). These are compared and contrasted with LQC in Section 5 of \cite{ag3}. Finally power suppression at large angular scales has been studied in the context of other bouncing models. In those discussions the bounce is often just assumed (as in \cite{cai1}), or obtained by adding a scalar field with a negative kinetic term which violates standard energy conditions ( as in \cite{cai2}). Our approach is different in that: (i) the bounce is a \emph{prediction} of LQC; (ii) since the mechanism has its roots in quantum geometry effects underlying LQG, additional scalar fields or violations of energy conditions are not involved; (iii) the standard inflationary potentials are used without adjustments to provide a fast roll phase; and, most importantly, (iv) our goal is to investigate whether the CMB observations can inform quantum gravity and vice versa.

In our view, the big bounce and the pre-inflationary dynamics of cosmological perturbations are on a robust footing although the discussion would benefit from a further sharpening of the detailed arguments that led us to the dressed metric $\t{g}_{ab}$. The part of the analysis that is on a less solid footing concerns the specific principles \cite{ag3,ag2} that were used to select the wave function $\Psi_o(a,\phi)$ of the background quantum geometry and the state $\psi(a, \phi)$ of perturbations. Note that these choices are necessary to make predictions in any approach that starts in the Planck regime. Indeed, even in standard inflation one has to assume that the perturbations are in the BD vacuum at the start of slow roll, and as discussed in section \ref{s3.3}, it is difficult to justify this assumption from first principles.  The fact that the principles led us to predictions that not only reproduce the successes of standard inflation, but also alleviate the tension associated with two anomalies, is an indication that they set us on the right track. However, they should be regarded as tentative first steps, to be improved upon and sharpened in future. In particular, the effect of the duration of the reheating epoch is yet to be investigated in detail. Another direction for future work is suggested by a second approach that has been used to alleviate  CMB anomalies using pre-inflationary dynamics \cite{sab,aks}. The point of departure is the same as in our approach: the bounce sets a new scale and modes with the longer wavelength at the bounce get excited during the pre-inflationary evolution and are no longer in the BD vacuum at the onset of inflation. However, the primordial spectrum does not exhibit power suppression (\ref{lqc}) as in our case. Instead, a key role is played by the superhorizon modes and their non-Gaussian correlations with the longest wavelength modes with those that are observable in the CMB. These correlations enhance the probability of finding certain features in the individual realizations of the primordial probability distribution, thereby alleviating anomalies associated with dipolar asymmetry and power suppression anomalies. It would be of considerable interest to re-examine all CMB anomalies from a perspective that combines strengths of the two approaches. Finally, all approaches to finding observable consequences of LQC in CMB, that we are aware of, assume an inflationary potential. At a certain level of discussion this is justified since a detailed analysis of the criticisms of the inflationary paradigm has concluded that, although important questions remain, the case for inflation has been ``strengthened by the PLANCK data" \cite{inflation-martin}. However, from a fundamental perspective, introduction of an inflaton and a specific potential is ad-hoc. Now, in full LQG, the Einstein-Hilbert Lagrangian receives higher order (in particular $R^2$) corrections because the curvature operator is constructed from Planck scale Wilson loops and hence non-local at the fundamental microscopic scale. Therefore it is important to analyze if these corrections would provide a natural basis for inflation purely from gravitational considerations, as in Starobinsky inflation.

\section*{Acknowledgments:} We are grateful to Donghui Jeong for numerous discussions on observational aspects discussed in this paper and for collaboration in \cite{agjs}. We would like to thank Ivan Agullo, Francois Bouchet, Woiciech Kaminski, Jerzy Lewandowski, Charles Lawrence, Jerome Martin and Patrick Peter for valuable discussions and Pawel Bielewicz  for help with Fig. \ref{fig:Ctheta}. This work was supported in part by the NSF grants PHY-1505411 and PHY-1806356,\, and the Eberly research funds of Penn State. Portions of this research were conducted with high performance computing resources provided by Louisiana State University (http://www.hpc.lsu.edu).

\end{document}